\documentclass[12pt]{article}

\usepackage{epsfig}

\def\Fint{\rlap{$\Biggl\rfloor$}\Biggl\lceil}
\def\square{\kern1pt\vbox{\hrule height 1.2pt\hbox{\vrule width 1.2pt\hskip 3pt
   \vbox{\vskip 6pt}\hskip 3pt\vrule width 0.6pt}\hrule height 0.6pt}\kern1pt}
\def\gtwid{\mathrel{\raise.3ex\hbox{$>$\kern-.75em\lower1ex\hbox{$\sim$}}}}
\def\ltwid{\mathrel{\raise.3ex\hbox{$<$\kern-.75em\lower1ex\hbox{$\sim$}}}}

\begin{document}

\begin{titlepage}
 
\begin{flushright}
SPIN-07/21, ITP-UU-07/31 \\ CRETE-06-13 \\ UFIFT-QG-06-05
\end{flushright}

\begin{center}
{\bf STOCHASTIC INFLATIONARY SCALAR ELECTRODYNAMICS}
\end{center}

\begin{center}
T. Prokopec$^*$
\end{center}

\begin{center}
\it{Institute for Theoretical Physics \& Spinoza Institute, Utrecht University\\
Leuvenlaan 4, Postbus 80.195, 3508 TD Utrecht, THE NETHERLANDS}
\end{center}

\begin{center}
N. C. Tsamis$^{\dagger}$
\end{center}

\begin{center}
\it{Department of Physics, University of Crete \\
GR-710 03 Heraklion, HELLAS}
\end{center}

\begin{center}
R. P. Woodard$^{\ddagger}$
\end{center}

\begin{center}
\it{Department of Physics, University of Florida \\
Gainesville, FL 32611, UNITED STATES}
\end{center}

\vspace{1cm}

\begin{center}
ABSTRACT
\end{center}
We stochastically formulate the theory of scalar quantum electrodynamics
on a de Sitter background. This reproduces the leading infrared logarithms 
at each loop order. It also allows one to sum the series of leading 
infrared logarithms to obtain explicit, nonperturbative results about
the late time behavior of the system. One consequence is confirmation of 
the conjecture by Davis, Dimopoulos, Prokopec and T\"ornkvist that 
super-horizon photons acquire mass during inflation. We compute $M^2_{\gamma}
\simeq 3.2991 \times H^2$. The scalar stays perturbatively light with 
$M^2_{\varphi} \simeq 0.8961 \times 3 e^2 H^2/8\pi^2$. Interestingly, the 
induced change in the cosmological constant is negative, $\delta \Lambda 
\simeq -0.6551 \times 3 G H^4/\pi$.

\begin{flushleft}
\end{flushleft}

\begin{flushleft}
$^*$ e-mail: T.Prokopec@phys.uu.nl \\
$^{\dagger}$ e-mail: tsamis@physics.uoc.gr \\
$^{\ddagger}$ e-mail: woodard@phys.ufl.edu
\end{flushleft}

\end{titlepage}

\section{Introduction}

Gravitons and massless, minimally coupled (MMC) scalars are unique in
being massless without classical conformal invariance. The combination
of these properties causes the accelerated expansion of spacetime 
during inflation to tear long wavelength virtual quanta out of the 
vacuum \cite{TW3,TW5,RPW1}. As more and more gravitons and MMC scalars emerge 
from the vacuum, the metric and MMC scalar field strengths experience 
a slow growth. The effect can be felt by any quantum field theory which 
involves either the undifferentiated metric or an undifferentiated MMC 
scalar.

A typical example is afforded by the MMC scalar with a quartic 
self-interaction,
\begin{equation}
\mathcal{L} = -\frac12 \Bigl(1 + \delta Z\Bigr) \partial_{\mu} \varphi 
\partial_{\nu} \varphi g^{\mu\nu} \sqrt{-g} - \frac12 \delta \xi \varphi^2
R \sqrt{-g} - \frac1{4!} \Bigl(\lambda + \delta \lambda\Bigr) \varphi^4 
\sqrt{-g} \; . \label{phi4}
\end{equation}
Consider this theory quantized on a nondynamical, locally de Sitter 
background,
\begin{equation}
ds^2 =-dt^2 + a^2 d\vec{x} \cdot d\vec{x} \qquad {\rm where} \qquad
a(t) = e^{H t} \; .
\end{equation}
If the finite parts of the renormalization constants (and the cosmological 
counterterm) are chosen to make the expectation value of the stress tensor 
vanish at $t=0$, then an explicit two loop computation using dimensional 
regularization reveals the following results for the induced energy density 
$\rho(t)$ and pressure $p(t)$ \cite{OW1,OW2},
\begin{eqnarray}
\rho(t) & = & \frac{\lambda H^4}{(2 \pi)^4} \Biggl\{ \frac18 \ln^2(a) 
\Biggr\} + O(\lambda^2) \; , \label{rho} \\
p(t) & = & \frac{\lambda H^4}{(2 \pi)^4} \Biggl\{ -\frac18 \ln^2(a) -
\frac1{12} \ln(a) \Biggr\} + O(\lambda^2) \; . \label{pres}
\end{eqnarray}

The factors of $\ln(a) = H t$ in expressions (\ref{rho}-\ref{pres}) are
known as {\it infrared logarithms}. They derive from the slow growth of
the scalar field amplitude that is apparent even in the free theory
\cite{VF,L,S},
\begin{equation}
\Bigl\langle \Omega_0 \Bigl\vert \varphi^2(x) \Bigr\vert \Omega_0 
\Bigr\rangle = {\rm Divergent\ Constant} + \frac{H^2}{(2\pi)^2} \ln(a) \; .
\end{equation}
The two loop expectation value of the stress tensor acquires two such factors
coming from the $-\frac1{4!} \lambda \varphi^4 g_{\mu\nu}$ term. 

Any quantum field theory which involves undifferentiated MMC scalars or 
metrics will show similar infrared logarithms in some of its Green's 
functions. They arise at one and two loop orders in the scalar 
self-mass-squared of this same theory \cite{BOW,KO}. In scalar quantum 
electrodynamics they have been seen in the one loop vacuum polarization 
\cite{PTW,PW1} and the two loop expectation values of certain scalar 
bilinears \cite{PTsW}. In Yukawa theory they show up in the one loop 
fermion self-energy \cite{PW2,GP} and in the two loop coincident vertex 
function \cite{MW1}. In pure quantum gravity they occur in the one 
loop graviton self-energy \cite{TW0} and in the two loop expectation value 
of the metric \cite{TW1}. When quantum gravity is coupled to a massless, 
Dirac fermion they occur in the one loop fermion self-energy \cite{MW2,MW3}. 
They even contaminate loop corrections to the power spectrum of cosmological 
perturbations \cite{SW1,BSV1,BSV2,MS,KC,BP} and other fixed-momentum 
correlators \cite{SW2}.

Infrared logarithms introduce a fascinating secular element into the usual, 
static results of quantum field theory. For example, without infrared 
logarithms, the expectation value of the stress tensor of (\ref{phi4}) would
be a constant times $g_{\mu\nu}$. With the same renormalization conventions 
\cite{OW1,OW2} in computing (\ref{rho}-\ref{pres}), the constant would 
actually be zero!

The most intriguing property of infrared logarithms is their ability to
compensate for powers of the loop counting parameter which suppress quantum 
loop effects. Indeed, the continued growth of $\ln(a) = H t$ must eventually 
{\it overwhelm} the loop counting parameter, no matter how small it is. 
However, this does not necessarily mean that quantum loop effects become 
strong. The correct conclusion is rather that perturbation theory breaks 
down past a given point in time. One must employ a nonperturbative technique 
to follow what happens at later times.

Certain models lend themselves to resummation schemes such as the $1/N$
expansion \cite{CM,BCVHSS} but a more general technique is suggested by the 
form of the expansion for $\rho(t)$ in (\ref{phi4}),
\begin{equation}
\rho(t) = H^4 \sum_{\ell=2}^{\infty} \lambda^{\ell-1} \Biggl\{ c_{\ell,0}
\Bigl[\ln(a)\Big]^{2\ell-2} + c_{\ell,1} \Bigl[\ln(a)\Bigr]^{2\ell-3} +
\dots + c_{\ell,2\ell-2} \ln^2(a)\Biggr\} . \label{genform}
\end{equation}
Here the constants $c_{\ell,k}$ are pure numbers which are assumed to be
of order one. The term in (\ref{genform}) involving $[\lambda \ln^2(a)]^{
\ell-1}$ is the {\it leading logarithm} contribution at $\ell$ loop order; 
the other terms are {\it subdominant logarithms}. Perturbation theory breaks 
down when $\ln(a) \sim 1/\sqrt{\lambda}$, at which point the leading 
infrared logarithms at each loop order contribute numbers of order one 
times $H^4$. In contrast, the subleading logarithms are all suppressed by
at least one factor of the small parameter $\sqrt{\lambda} \ll 1$. So it 
makes sense to retain only the leading infrared logarithms,
\begin{equation}
\rho(t) \longrightarrow H^4 \sum_{\ell=2}^{\infty} c_{\ell,0} \Bigl[\lambda
\ln^2(a)\Bigr]^{\ell-1} \; .
\end{equation}
This is known as the {\it leading logarithm approximation}.

Starobinski\u{\i} has developed a simple stochastic formalism \cite{AAS}
which reproduces the leading infrared logarithms at each order for any 
scalar potential model of the form,
\begin{equation}
\mathcal{L} = -\frac12 (1 + \delta Z) \partial_{\mu} \varphi \partial_{\nu} 
\varphi g^{\mu\nu} \sqrt{-g} - V(\varphi) \sqrt{-g} \; . \label{Starform}
\end{equation}
Probabilistic representations of inflationary cosmology have been much
studied in order to understand initial conditions \cite{AV,NS} and global 
structure \cite{GLM,LM}. However, we wish here to focus on Starobinski\u{\i}'s
technique as a wonderfully simple way of recovering the most important
secular effects of inflationary quantum field theory \cite{SJR,SNN,WV,MM}. 
It is of particular importance for us that Starobinski\u{\i} and Yokoyama 
have shown how to take the late time limit of the series of leading 
infrared logarithms whenever the potential $V(\varphi)$ is bounded below 
\cite{SY}. This is the true analogue of what the renormalization group
accomplishes in flat space quantum field theory and statistical mechanics.

The solution of Starobinski\u{\i} and Yokoyama is an amazing achievement,
but it only gives us nonperturbative control over the infrared logarithms 
which arise in scalar potential models (\ref{Starform}). The most general 
models which show infrared logarithms possess two complicating features:
\begin{itemize}
\item{Couplings to fields other than MMC scalars and gravitons; and}
\item{Interactions which involve differentiated MMC scalars and 
gravitons.\footnote{Of course there would be no infrared logarithms if 
{\it all} the MMC scalars and gravitons were differentiated. However, 
infrared logarithms must arise, in the expectation values of some operators, 
from interactions which involve at least one undifferentiated MMC scalar or 
graviton. Examples include the $h^n \partial h \partial h$ interaction of 
pure quantum gravity \cite{TW0,TW1} and scalar interactions of the form 
$\varphi^2 \partial \varphi \partial \varphi$ \cite{SW1,TW2}.}}
\end{itemize}
An important step forward was a recent leading log solution for the model
comprised by a MMC scalar which is Yukawa-coupled to a massless, Dirac fermion
\cite{MW1}. That model possesses the first complicating feature but not the
second. In this paper we derive a similar leading log solution for MMC 
scalar quantum electrodynamics (SQED),
\begin{eqnarray}
{\cal L} &\! = \!&
- (1 + \delta Z_2) \, (\partial_{\mu} - ieA_{\mu}) \phi^{*} \,
(\partial_{\nu} + ieA_{\nu}) \phi \; g^{\mu\nu} \sqrt{-g}
- \; \delta\xi \, \phi^{*} \phi \; R \, \sqrt{-g} \;
\nonumber \\
& \mbox{} &
- \, \frac14 (1 + \delta Z_3) \, F_{\mu\nu} \, F_{\rho\sigma} \,
g^{\mu\rho} \, g^{\nu\sigma} \sqrt{-g} \;
- \, \frac{\delta\lambda}{4} \, (\phi^{*} \phi)^2 \, \sqrt{-g}
\;\; , \label{L}
\end{eqnarray}
Although this model has derivative interactions we will see that a
gauge choice permits one to avoid them at leading log order. We still
do not have a full understanding of how to treat derivative interactions.

Section 2 of this paper summarizes a recent all-order derivation 
\cite{RPW2,TW2} of the Starobinski\u{\i} formalism for scalar potential 
models. Of course it is crucial to understand why the technique works in 
order to apply it to more general theories. That problem is discussed in 
section 3, reaching the conclusion that one deals with other fields by 
integrating them out and then stochastically simplifying the resulting 
effective action. Section 4 accomplishes this for the vector potential 
of SQED. One must also integrate out the vector potential from any operator 
whose expectation value is desired. The resulting, purely scalar operator 
is then stochastically simplified before computing the leading log 
contribution to its expectation value using Starobinski\u{\i}'s technique. 
We do this in section 5 for the various constituents of the SQED stress 
tensor. Because the leading logarithm limit of SQED gives a model of the 
form (\ref{Starform}), with a potential which is bounded below, one can 
exploit the solution of Starobinski\u{\i} and Yokoyama to make explicit, 
nonperturbative predictions. We do this in section 6 for the expectation 
values of $\varphi^*(x) \varphi(x)$, $F_{\mu\nu}(x) F_{\rho\sigma}(x)$ 
and $T_{\mu\nu}(x)$. We also confirm the remarkable conjecture of Davis, 
Dimopoulos, Prokopec and Tornkvist that super-horizon photons acquire 
mass during inflation \cite{DDPT,DPTD}. The eventual, nonperturbative 
photon mass-squared turns out to be about a hundred times larger than 
perturbative estimates.

\section{Deriving Starobinski\u{\i}'s Formalism}

Infrared logarithms arise in explicit perturbative computations of
Green's functions formed from quantum field operators. It is an amazing 
fact that one can reproduce the leading infrared logarithms in any model 
of the form (\ref{Starform}) using a formalism in which the fields are 
classical random variables. One wonders, what became of the Uncertainty 
Principle? What became of the ultraviolet divergences and the counterterms?
And how did the stochastic jitter emerge?

In order to apply Starobinski\u{\i}'s formalism to more general models 
one must understand why it works. That is the task of this section. We
begin by expressing the dimensionally regulated, Heisenberg field 
equations in Yang-Feldman form \cite{YF}. We then explain how infrared
logarithms arise and, of crucial importance, the conditions for any
expectation value of undifferentiated fields to receive a leading logarithm 
contribution. Based upon this understanding, we identify a series of 
simplifications that can be made to the Yang-Feldman equation without in 
any way affecting the leading infrared logarithms. Taking the time 
derivative of this simplified Yang-Feldman equation results in 
Starobinski\u{\i}'s Langevin equation, with the white noise emerging as 
the time derivative of the simplified free field. The section closes with 
a review of the nonperturbative solution of Starobinski\u{\i} and Yokoyama 
obtained \cite{SY} for the late time limit of any model with a potential 
which is bounded below.

\subsection{The Free Field Expansion}

The dimensionally regulated, Heisenberg operator equation for 
(\ref{Starform}) is,
\begin{equation}
\ddot{\varphi} + (D-1) H \dot{\varphi} - \frac{\nabla^2}{a^2} \varphi + 
\frac{V'(\varphi)}{1 + \delta Z} = 0 \; .
\end{equation}
Integrating it results in the Yang-Feldman equation \cite{YF}, 
\begin{equation}
\varphi(t,\vec{x}) = \varphi_0(t,\vec{x}) - \int_0^t dt' \, a^{\prime D-1} 
\int d^{D-1}x' \, G(x;x') \frac{V'\Bigl(\varphi(x')\Bigr)}{1 + \delta Z} \; .
\label{YFeqn}
\end{equation}
A number of quantities in (\ref{YFeqn}) require definition. The {\it free 
field} $\varphi_0(x)$ and the {\it retarded Green's function} $G(x;x')$ 
are,
\begin{eqnarray}
\varphi_0(t,\vec{x}) &\!\!\equiv\!\! & \int \!\! \frac{d^{D-1}k}{(2\pi)^{D-1}} 
\, \theta(k - H) \Bigl\{ u(t,k) e^{i \vec{k} \cdot \vec{x}} \alpha(\vec{k}) + 
u^*(t,k) e^{-i \vec{k} \cdot \vec{x}} \alpha^{\dagger}(\vec{k}) \Bigr\} 
\label{phi0} \; , \qquad \\
G(x;x') & \!\!\equiv \!\!& i\theta(\Delta t) \!\! \int \!\! \frac{d^{D-1}k}{(2
\pi)^{D-1}} \, e^{i \vec{k} \cdot \Delta \vec{x}} \Bigl\{ u(t,k) u^*(t',k) 
- u^*(t,k) u(t',k) \Bigr\} \; . \qquad \label{Gret}
\end{eqnarray}
Here $\Delta t \equiv t - t'$, $\Delta \vec{x} \equiv \vec{x} - \vec{x}'$ 
and the mode function $u(t,k)$ is,
\begin{equation}
u(t,k) = i \sqrt{\frac{\pi}{4 H a^{D-1}}} 
H^{(1)}_{\frac{D-1}2}\Bigl(\frac{k}{H a}\Bigr) = \frac{\Gamma(\frac{D-1}2)}{
\sqrt{4 \pi H}} \Bigl(\frac{2 H}{k}\Bigr)^{\frac{D-1}2} \Biggl\{1 +
O\Bigl(\frac{k^2}{H^2 a^2}\Bigr) \Biggr\} . \label{uform}
\end{equation}
The nonzero commutation relations of the canonically normalized creation 
and annihilation operators are,
\begin{equation}
\Bigl[\alpha(\vec{k}), \alpha^{\dagger}(\vec{k}')\Bigr] = (2\pi)^{D-1}
\delta^{D-1}(\vec{k} - \vec{k}') \; .
\end{equation}
It follows that the retarded Green's function can be expressed as the 
commutator of two free fields,\footnote{The restriction to $k \equiv 
\Vert \vec{k} \Vert \geq H$ in 
the free field mode sum (\ref{phi0}) is imposed to avoid an infrared 
singularity in the free propagator \cite{FP}. The physical reason for this 
singularity is that no causal process would allow an experimenter to 
prepare the initial state in coherent Bunch-Davies vacuum over an infinite 
spatial section. Sensible physics can be regained either by employing an 
initial state for which the super-horizon modes are less strongly 
correlated \cite{AV2}, or else by working on a compact spatial manifold 
such as $T^{D-1}$ for which there are initially no super-horizon modes 
\cite{TW3}. In both cases the modes with $k < H$ are effectively absent. 
Note also that one typically removes the cutoff on any mode sum, such as 
(\ref{Gret}), which is not singular at $k=0$.}
\begin{equation}
G(x;x') = i \theta(\Delta t) \Bigl[\varphi_0(x),\varphi_0(x')\Bigr] \; .
\end{equation}

Iterating the Yang-Feldman equation generates the usual interaction
picture expansion of the field, in this case expressed in terms of 
a free field $\varphi_0(x)$ which agrees with the full field and its 
first time derivative at $t=0$. Without worrying about operator ordering, 
we can write out the first few terms of this expansion,
\begin{eqnarray}
\lefteqn{\varphi(x) = \varphi_0(x) - \int d^Dx' \sqrt{-g(x')} \, G(x;x')
\frac{V'\Bigl(\varphi_0(x')\Bigr)}{1 + \delta Z} } \nonumber \\
& & \hspace{2.5cm} + \int d^Dx' \sqrt{-g(x')} \, G(x;x') \frac{V''\Bigl(
\varphi_0(x')\Bigr)}{1 + \delta Z} \nonumber \\
& & \hspace{3.5cm} \times \int d^Dx'' \sqrt{-g(x'')} \, G(x';x'') 
\frac{V'\Bigl(\varphi_0(x'')\Bigr)}{1 + \delta Z} + \ldots \; . \qquad
\end{eqnarray}
A nice diagrammatic representation for this expansion has recently been 
given by Musso \cite{Mus}.

The integrals over $x^{\prime \mu}$ and $x^{\prime \prime \mu}$ are
known as {\it vertex integrations}. The expectation value of any operator
which involves $\varphi(x)$ --- for example, $\varphi^N(x)$ or $\varphi(x)
\varphi(x')$ --- can obviously be reduced to a sum of terms, each one of 
which consists of a number of vertex integrations of Green's functions
times the expectation value of some number of free fields. The expectation
value of free fields can be further reduced to a sum of products of 
expectation values of two free fields. 

\subsection{Genesis of Infrared Logarithms}

Infrared logarithms derive from two sources: the expectation values of 
pairs of free fields and the vertex integrations.

$\bullet$ {\it Free Field Expectation Values:} Let us first consider the 
expectation value of two free fields in the presence of the state which 
obeys $\alpha(\vec{k}) \vert \Omega \rangle = 0$,
\begin{eqnarray}
\Bigl\langle \Omega \Bigl\vert \varphi_0(x) \varphi_0(x')
\Bigr\vert \Omega \Bigr\rangle & = & \int \frac{d^{D-1}k}{(2\pi)^{D-1}} 
\, \theta(k - H) e^{i \vec{k} \cdot \Delta \vec{x}} u(t,k) u^*(t',k) 
\; , \qquad \\
& = & \int_H^{\infty} dk \, k^{D-2} \frac{J_{\frac{D-3}2}(k \Delta x) u(t,k) 
u^*(t',k)}{2^{D-2} \pi^{\frac{D-1}2} (\frac{k \Delta x}2)^{\frac{D-3}2}}
\; . \qquad \label{modesum}
\end{eqnarray}
At high $k$ the Bessel functions in (\ref{modesum}) oscillate for $x^{\mu} 
\neq x^{\prime \mu}$, which makes the integral converge. Even at coincidence 
there is no possibility of an $a$-dependent ultraviolet divergence,
\begin{eqnarray}
\Bigl\langle \Omega \Bigl\vert \varphi^2_0(x) \Bigr\vert \Omega \Bigr\rangle 
& = & \frac1{2^{D} \pi^{\frac{D-3}2} \Gamma(\frac{D-1}2) H a^{D-1}} 
\int_H^{\infty} dk \, k^{D-2} \Bigl\Vert H^{(1)}_{\frac{D-1}2}\Bigl(
\frac{k}{H a}\Bigr) \Bigr\Vert^2 \; , \qquad \\
& = & \frac{H^{D-2}}{2^{D} \pi^{\frac{D-3}2} \Gamma(\frac{D-1}2)} 
\int_{\frac1{a}}^{\infty} dz \, z^{D-2} \Bigl\Vert H^{(1)}_{\frac{D-1}2}(z)
\Bigr \Vert^2 \; . \qquad
\end{eqnarray}
The only possible $a$ dependence is the finite contribution from the 
infrared.

The small $k$ expansion of the integrand in (\ref{modesum}) is,
\begin{eqnarray}
\lefteqn{k^{D-2} \frac{J_{\frac{D-3}2}(k \Delta x) u(t,k) u^*(t',k)}{2^{D-2} 
\pi^{\frac{D-1}2} (\frac{k \Delta x}2)^{\frac{D-3}2}} } \nonumber \\
& & \hspace{3cm} = \frac{\Gamma(\frac{D-1}2) H^{D-2}}{2 \pi^{\frac{D+1}2} k} 
\Biggl\{1 + O\Bigl(\frac{k^2}{H^2 a^2},\frac{k^2}{H^2 a^{\prime 2}},k^2 
\Delta x^2\Bigr) \Biggr\} . \label{lowk} \qquad \label{smallk}
\end{eqnarray}
Had the lower limit not been cut off at $k = H$ this would give a logarithmic
divergence. With the infrared cutoff there is no divergence, but one does get
a large logarithm. It derives exclusively from the first term of (\ref{lowk}),
integrated up to the point where the expansion breaks down and the integrand
begins to oscillate. This point is $k \simeq {\rm Min}(Ha,Ha',1/\Delta
x)$. By taking account of causality,
\begin{equation}
\Delta x \leq \Bigl\vert \frac1{H a} - \frac1{Ha'} \Bigr\vert \qquad 
\Longrightarrow \qquad \frac1{\Delta x} \geq {\rm Min}(Ha,Ha') \; ,
\end{equation}
we see that the upper limit is actually $k = {\rm Min}(Ha,Ha') \equiv 
\alpha$,\footnote{We have also exploited the doubling formula \cite{GR}
to write,
\begin{eqnarray}
\Gamma\Bigl(\frac{D\!-\!1}2\Bigr) = \frac{\sqrt{\pi}}{2^{D-2}} \frac{\Gamma(D
\!-\! 1)}{\Gamma(\frac{D}2)} \; . \nonumber
\end{eqnarray}}
\begin{equation}
\frac{\Gamma(\frac{D-1}2) H^{D-2}}{2 \pi^{\frac{D+1}2}} \int_{H}^{H \alpha} 
\frac{dk}{k} = \frac{H^{D-2}}{(4\pi)^{\frac{D}2}} \frac{\Gamma(D\!-\!1)}{
\Gamma(\frac{D}2)} 2 \ln(\alpha) \longrightarrow \frac{H^2}{4 \pi^2} 
\ln(\alpha) \quad ({\rm in} \; D=4) \; . \label{arbD}
\end{equation}
To summarize, {\it infrared logarithms from $\varphi_0(t,\vec{x})$ derive 
exclusively from the range $H \ltwid k \ltwid H a(t)$. Further, only the
first term in the long wavelength expansion (\ref{uform}) of the mode 
functions contributes.}

$\bullet$ {\it Vertex Integrations:} We turn now to the second source of 
infrared logarithms, which is vertex integrations. It is apparent from 
expressions (\ref{Gret}-\ref{uform}) that the most infrared-singular part 
of the mode function drops out of the retarded Green's function, 
\begin{eqnarray}
\lefteqn{i\theta(\Delta t) \Bigl[u(t,k) u^*(t',k) - u^*(t,k) u(t',k)\Bigr] }
\nonumber \\
& & \hspace{2.5cm} = \frac{\theta(\Delta t)}{(D-1) H} \Bigl[\frac1{a^{\prime 
D-1}} - \frac1{a^{D-1}}\Bigr] \Biggl\{ 1 + O\Bigl(\frac{k^2}{H^2 a^2},
\frac{k^2}{H^2 a^{\prime 2}}\Bigr)\Biggr\} . \qquad \label{Gexp}
\end{eqnarray}
Hence the retarded Green's function cannot contribute infrared logarithms.
However, consider the vertex integration of $G(x;x')$ against $n$ powers of
$\ln(a')$,
\begin{eqnarray}
\lefteqn{\int_0^t dt' \, a^{\prime D-1} \int d^{D-1}x' \, G(x;x') 
\Bigl[\ln(a')\Bigr]^n } \nonumber \\
& & \hspace{3cm} = \frac1{(D-1) H} \int_0^t dt' \, \Bigl[1 - \Bigl(\frac{a'}{a}
\Bigr)^{D-1} \Bigr] \Bigl[\ln(a')\Bigr]^n \; , \label{intermediate} \qquad \\
& & \hspace{3cm} = \frac1{(n+1) (D-1) H^2} \Biggl\{ \Bigl[\ln(a)\Bigr]^{n+1} + 
O\Bigl([\ln(a)]^n\Bigr) \Biggr\} . \label{final} \qquad 
\end{eqnarray}
The temporal vertex integration has increased the number of infrared 
logarithms from $n$ to $n+1$.

A temporal vertex integration can only produce an additional infrared
logarithm when it receives nearly equally weighted contributions from
its full range. This requires that no factors of $a'$ should remain 
after multiplying by the $a^{\prime D-1}$ from the measure and performing
the spatial vertex integrations. For example, consider the two terms from 
the right hand side of (\ref{intermediate}),
\begin{eqnarray}
\int_0^t dt' \, [\ln(a')]^n & = & \int_0^t dt' \, (Ht')^n = 
\frac{[\ln(a)]^{n+1}}{(n+1) H} \; , \qquad \\
\int_0^t dt' \, \Bigl(\frac{a'}{a}\Bigr)^{D-1} [\ln(a')]^n & = & 
e^{-\beta H t} \Bigl(\frac{\partial}{\partial \beta}\Bigr)^n \Biggl[
\frac{e^{\beta H t} - 1}{\beta H} \Biggr]_{\beta = D-1} \; , \\
& = & \frac{[\ln(a)]^n}{(D-1) H} - \frac{n [\ln(a)]^{n-1}}{(D-1)^2 H} + 
\ldots \; . \qquad
\end{eqnarray}
To summarize, {\it infrared logarithms from vertex integrations derive 
entirely from the part of the long wavelength expansion of the Green's 
function which goes like $1/a^{\prime D-1}$.}

\subsection{Conditions for a Leading Log Contribution}

We have seen the ways in which infrared logarithms originate from 
the expectation values of pairs of free fields, and from vertex 
integrations. It is now time to consider the crucial issue of how 
many infrared logarithms must derive from each source in order to 
reach leading log order. Although the answer is completely general,
it is easier to explain in the context of the quartic self-interaction,
\begin{equation}
V(\varphi) = \frac12 \delta \xi \varphi^2 R + \frac1{4!} (\lambda + 
\delta \lambda) \varphi^4 \; .
\end{equation}

Let us first establish that the various counterterms cannot make leading
log contributions. This follows from the number of fields they carry and
the number of factors of $\lambda$ they involve in this model 
\cite{OW1,OW2,BOW},
\begin{equation}
\delta Z = O(\lambda) \qquad , \qquad \delta \xi = O(\lambda) \qquad ,
\qquad \delta \lambda = O(\lambda^2) \; .
\end{equation}
The coupling constant renormalization, $\delta \lambda$, multiplies the
same $\varphi^4$ term as $\lambda$, so contributions involving it produce 
the same structure of infrared logarithms as contributions from $\lambda
\varphi^4$. Because contributions from $\delta \lambda \varphi^4$ have
at least one extra factor of $\lambda$, with no more infrared logarithms,
they can never be leading order. The same argument applies to the field
strength renormalization, $\delta Z$, on account of the fact that it enters
the field equations in the form, $V'(\varphi)/(1 + \delta Z)$. While both
$\delta \xi$ and $\lambda$ go like $\lambda$, the former multiplies 
$\varphi^2$ while the latter multiplies $\varphi^4$. So no contribution 
involving $\delta \xi$ can produce as many infrared logarithms, at the same 
order in $\lambda$, as contributions involving only the $\lambda \varphi^4$
term. We therefore get exactly the same leading log contributions from the 
simplified Yang-Feldman equation without the counterterms,
\begin{equation}
\varphi(x) = \varphi_0(x) - \frac{\lambda}6 \int_0^t dt' \, a^{\prime D-1} 
\int d^{D-1} x' \, G(x;x') \varphi^3(x') \; . \label{simple}
\end{equation}

Let us consider the generic form of the free field expansion derived from 
(\ref{simple}), keeping track only of the number of $\lambda$'s, the
number of vertex integrations, and the number of fields at any point.
We already have symbols for the coupling constant $\lambda$ and the field
$\varphi_0$. Let us employ the symbol ``$I$'' to denote generic vertex 
integrations,
\begin{equation}
I \equiv \int_0^t dt' \, a^{\prime D-1} \int d^{D-1} x' G(x;x') \; .
\end{equation}
In this notation we might render (\ref{simple}) as follows,
\begin{equation}
\varphi \sim \varphi_0 + \lambda I \varphi^3 \; . \label{verysimple}
\end{equation}
Note that we do not worry about signs or numerical factors such as $1/6$,
nor do we worry about which spacetime points the various fields reside.
In this generic language it is simple to iterate (\ref{verysimple}) to
exhibit the generic form of the free field expansion,
\begin{equation}
\varphi \sim \varphi_0 + \lambda I \varphi_0^3 + \lambda^2 I^2 \varphi_0^5
+ \lambda^3 I^3 \varphi_0^7 + \ldots \; . \label{generic}
\end{equation}
In other words, {\it each additional factor of $\lambda$ involves one more
vertex integration and two more free fields.}

The same generic form (\ref{generic}) applies to any operator whose VEV
we might wish to compute. For example, the product of $N$ $\varphi$'s ---
even at all different points --- would be rendered,
\begin{equation}
\varphi^N \sim (\varphi_0)^N \Bigl\{ 1 + \lambda I \varphi_0^2 + \lambda^2
I^2 \varphi_0^4 + \lambda^3 I^3 \varphi_0^6 + \ldots \Bigr\} . \label{genop}
\end{equation}
Note that we do not worry, at this level, about which spacetime points the 
various free fields reside, or which of the parent full fields gives rise
to corrections in the free field expansion. The only interesting things
are the number of factors of $\lambda$, the number of vertex integrations, 
and the number of free fields.

Now recall that leading log corrections to any result in this theory must
produce two infrared logarithms for each extra factor of the loop-counting
parameter $\lambda$ \cite{OW1,OW2,BOW,KO}. From the generic free field 
expansion (\ref{genop}) we see that each additional factor of $\lambda$
is accompanied by one new vertex integration and two new free fields.
Because a vertex integration can add at most a single infrared logarithm,
as can any pair of free fields, it follows that {\it leading log contributions 
to the expectation value of any combination of undifferentiated full fields 
$\varphi$ can only result when each vertex integration and each free field 
contributes to an infrared logarithm.} That turns out to be a general 
conclusion for any potential $V(\varphi)$.

\subsection{Starobinski\u{\i}'s Langevin Equation}

It is well to summarize what we have learned in the previous sub-sections:
\begin{enumerate}
\item{For the expectation value of any combination of undifferentiated full
fields to receive a leading log contribution, every $\langle \varphi_0(x) 
\varphi_0(x') \rangle$ and every vertex integration must contribute an 
infrared logarithm.}
\item{The infrared logarithm from $\varphi_0(t,\vec{x})$ derives exclusively
from modes in the range $H < k < H a(t)$, and from just the leading term in
the small $k$ expansion of $u(t,k)$. That is, the following replacement makes
no change at leading log order,
\begin{eqnarray}
\lefteqn{\varphi_0(t,\vec{x}) \longrightarrow \int 
\frac{d^{D-1}k}{(2\pi)^{D-1}} \, \theta(k - H) \theta(H a - k) } \nonumber \\
& & \hspace{2.8cm} \times \frac{\Gamma(\frac{D-1}2)}{\sqrt{4 \pi H}}
\Bigl(\frac{2 H}{k}\Bigr)^{\frac{D-1}2} \Bigl\{e^{i \vec{k} \cdot \vec{x}} 
\alpha(\vec{k}) + e^{-i \vec{k} \cdot \vec{x}} \alpha^{\dagger}(\vec{k})
\Bigr\} . \qquad \label{1stsimp}
\end{eqnarray}}
\item{The infrared logarithm from a vertex integration derives exclusively 
from the term in the small $k$ expansion of $u(t,k) u^*(t',k) - u^*(t,k) 
u(t',k)$ which goes like $1/a^{\prime D-1}$. That is, the following replacement
makes no change at leading log order,
\begin{equation}
G(x;x') \longrightarrow \frac{\theta(\Delta t) \delta^{D-1}(\vec{x} - 
\vec{x}')}{(D-1) H a^{\prime D-1}} \; . \label{2ndsimp}
\end{equation}}
\end{enumerate}

Because of point 1 we can make the simplifications 
(\ref{1stsimp}-\ref{2ndsimp}) in the Yang-Feldman equation. Because the
infrared truncation of (\ref{1stsimp}) removes any possibility for
ultraviolet divergences, we can set $D=4$ to write the infrared truncated
free field as,
\begin{equation}
\Phi_0(t,\vec{x}) = \int \frac{d^3k}{(2\pi)^3} \, \theta(k - H) \theta(H a - k) 
\frac{H}{\sqrt{2 k^3}} \Bigl\{e^{i \vec{k} \cdot \vec{x}} \alpha(\vec{k}) 
+ e^{-i \vec{k} \cdot \vec{x}} \alpha^{\dagger}(\vec{k}) \Bigr\} . 
\end{equation}
Note that we have used a new symbol, $\Phi_0(x)$, to distinguish it from
the original free field, $\varphi_0(x)$. Whereas $\varphi_0(x)$ embodies 
the Uncertainty Principle, the infrared truncated free field does not,
\begin{equation}
\Bigl[\Phi_0(x),\Phi_0(x')\Bigr] = 0 \; .
\end{equation}
Note also that, whereas the VEV of $\varphi^2_0(x)$ diverges, the VEV of
$\Phi^2_0(x)$ does not. We have proved that these very different quantities 
nevertheless produce precisely the same infrared logarithms.

We have already seen that counterterms can be dropped at leading log order.
The resulting, simplified Yang-Feldman equation is accordingly,
\begin{equation}
\Phi(t,\vec{x}) = \Phi_0(t,\vec{x}) - \frac1{3 H} \int_0^t dt' \,
V'\Bigl(\Phi(t',\vec{x})\Bigr) \; . \label{newYF}
\end{equation}
Note that we have called the field $\Phi(x)$ to distinguish it from 
$\varphi(x)$. This field $\Phi(x)$ commutes with $\Phi(x')$ for all 
$x^{\prime\mu}$, just like the free field, $\Phi_0(x)$. Further, VEV's 
involving it are competely free of ultraviolet divergences. They 
nevertheless agree exactly with VEV's of $\varphi(x)$ at leading log order.

Taking the time derivative of (\ref{newYF}) gives Starobinski\u{\i}'s 
Langevin equation \cite{SY},
\begin{equation}
\dot{\Phi}(t,\vec{x}) = \dot{\Phi}_0(t,\vec{x}) - \frac1{3 H} V'\Bigl(
\Phi(t,\vec{x})\Bigr) \; . \label{Langevin}
\end{equation}
Starobinski\u{\i}'s stochastic noise term is the time derivative of the
infrared truncated free field,
\begin{equation}
\dot{\Phi}_0(t,\vec{x}) = \int \frac{d^3k}{(2\pi)^3} \delta(Ha - k) 
\frac{H^2}{\sqrt{2 k}} \Bigl\{e^{i\vec{k} \cdot \vec{x}} \alpha(\vec{k}) +
e^{-i \vec{k} \cdot \vec{x}} \alpha^{\dagger}(\vec{k})\Bigr\} \; .
\end{equation}
A simple calculation reveals that it behaves like white noise,
\begin{equation}
\Bigl\langle \Omega \Bigl\vert \dot\Phi_0(t,\vec{x}) \dot\Phi_0(t',\vec{x}) 
\Bigr\vert \Omega \Bigr\rangle = \frac{H^3}{4 \pi^2} \delta(t - t') \; . 
\label{noise}
\end{equation}

\subsection{Nonperturbative Solution}

Langevin equations of the form (\ref{Langevin}) have been much studied 
\cite{ALV}. Expectation values of functionals of the stochastic field can 
be computed in terms of a probability density $\varrho(t,\phi)$ as follows,
\begin{equation}
\Big\langle \Omega \Bigl\vert F\left[ \Phi(t, {\vec x})\right]
\Bigr\vert \Omega \Big\rangle = \int_{-\infty}^{+\infty} d\phi \,
\varrho(t,\phi) F(\phi) \; .
\end{equation}
The probability density satisfies a Fokker-Planck equation whose first 
term is given by the interaction in (\ref{Langevin}) and whose second term 
is fixed by the normalization of the white noise (\ref{noise}):
\begin{equation}
\dot{\varrho}(t,\phi) = \frac{1}{3H} \frac{\partial}{\partial \phi}
\Bigl[ V'(\phi) \varrho(t,\phi) \Bigr] + \frac12 \frac{\partial^2}{\partial 
\phi^2} \Bigl[\frac{H^3}{4 \pi^2} \varrho(t,\phi) \Bigr] \; .
\end{equation}
To recover the nonperturbative late time solution of Starobinski\u{\i} 
and Yo\-ko\-ya\-ma \cite{SY} one makes the ansatz,
\begin{equation}
\lim_{t \rightarrow \infty} \varrho(t,\phi) = \varrho_{\infty}(\phi) \; , 
\end{equation}
because the $-V'(\varphi)$ force should eventually balance the tendency of
inflationary particle production to push the scalar up its potential.
This ansatz results in a first order equation,
\begin{equation}
\frac{d \varrho_{\infty}(\phi)} {\varrho_{\infty}(\phi)} =
- \frac{8 \pi^2}{3 H^4} V'(\phi) d\phi \; .
\end{equation}
The solution is straightforward,
\begin{equation}
\varrho_{\infty}(\phi) = N \exp\left[-\frac{8 \pi^2}{3 H^4} V(\phi)\right] \; .
\end{equation}

\section{Generalizing Starobinski\u{\i}'s Formalism}

In this section we discuss the problem of generalizing Starobinski\u{\i}'s
formalism beyond scalar potential models. We begin by explaining the two
complicating features: derivative interactions and passive fields which do 
not produce infrared logarithms. We then derive a formula which gives the
leading logarithm series for a general interaction. By applying this
formula to simple models it emerges that ultraviolet divergences can
contaminate even the leading infrared logarithms. This is because passive
fields --- and differentiated fields of any type --- make contributions
of order one which multiply the leading logarithms contributed by other
fields. Unlike the infrared logarithms, these order one contributions
derive from all portions of the free field mode sum, and from the full
free field mode functions. Hence it is not valid to stochastically
simplify passive fields as we did in the previous section. The correct
procedure is to integrate them out, and then stochastically simplify
the resulting effective field equations. We show that this amounts to 
computing the effective potential.

\subsection{Active, Passive and Differentiated Fields}

We use the term {\it active} to denote a field whose mode functions are 
right to produce infrared logarithms. Fields whose mode functions cannot
produce infrared logarithms are called {\it passive}. A typical passive
field is the massless, conformally coupled scalar,
\begin{equation}
\mathcal{L} = -\frac12 \partial_{\mu} \psi \partial_{\nu} \psi g^{\mu\nu}
\sqrt{-g} - \frac18 \Bigl(\frac{D-2}{D-1}\Bigr) \psi^2 R \sqrt{-g} \; .
\end{equation}
The plane wave mode functions for this field can be worked out for any 
scale factor,
\begin{equation}
v(t,k) = \frac{a^{1 - \frac{D}2}}{\sqrt{2 k}} \exp\Bigl[-i k \int_0^t 
\frac{dt'}{a(t')}\Bigr] \; .
\end{equation}
The shall henceforth make the specialization to de Sitter so that,
\begin{equation}
{\rm de \; Sitter} \qquad \Longrightarrow \qquad
v(t,k) = \frac{a^{1 - \frac{D}2}}{\sqrt{2 k}} \exp\Bigl[
-\frac{i k}{H} + \frac{ik}{Ha} \Bigr] \; .
\end{equation}

These mode functions are not singular enough for the expectation value of 
two free fields to produce an infrared logarithm for $D > 2$,
\begin{eqnarray}
\Bigl\langle \Omega \Bigl\vert \psi(x) \psi(x') \Bigr\vert \Omega \Bigr\rangle
& = & \int \frac{d^{D-1}k}{(2\pi)^{D-1}} \, e^{i \vec{k} \cdot \Delta \vec{x}}
v(t,k) v^*(t',k) \; , \\
& = & \frac{(a a')^{1 - \frac{D}2}}{(4 \pi)^{\frac{D-1}2}} \int_0^{\infty}
dk \, k^{D-3} \frac{J_{\frac{D-3}2}(k \Delta x) e^{\frac{ik}{H a} -
\frac{ik}{H a'}}}{(\frac{k \Delta x}2)^{\frac{D-3}2}} \; .
\end{eqnarray}
Nor, again for $D > 2$, can one get an infrared logarithm from the vertex 
integration of the retarded conformal Green's function,
\begin{equation}
\int_0^t dt' \, a^{\prime D-1} \int d^{D-1}x' \, G_{\rm cf}(x;x') = \frac1{H}
\int_0^t dt' \, \Bigl[ \Bigl(\frac{a'}{a}\Bigr)^{\frac{D}2-1} - 
\Bigl(\frac{a'}{a}\Bigr)^{\frac{D}2} \Bigr] \; .
\end{equation}

The absence of infrared logarithms from either source for the conformally
coupled scalar is obviously related because its mode functions obey the same
Wronskian as do the minimally coupled scalar mode functions,
\begin{equation}
v(t,k) \dot{v}^*(t,k) - \dot{v}(t,k) v^*(t,k) = \frac{i}{a^{D-1}} =
u(t,k) \dot{u}^*(t,k) - \dot{u}(t,k) u^*(t,k) \; .
\end{equation}
This relation requires that a total of $D-1$ factors of $a$ must be
shared between any two linearly independent solutions. For example, the
real and imaginary parts obey,
\begin{equation}
-2 {\rm Re}(v) {\rm Im}(\dot{v}) + 2 {\rm Re}(\dot{v}) {\rm Im}(v) = 
\frac1{a^{D-1}} = -2 {\rm Re}(u) {\rm Im}(\dot{u}) + 2 {\rm Re}(\dot{u}) 
{\rm Im}(u) \; .
\end{equation}
For the VEV of a pair of free fields to produce an infrared logarithm
requires that the far infrared ($\vec{k} \approx 0$) mode function should
approach a phase divided by $k^{\frac{D-1}2}$, with no dependence
upon $a$. If we make the phase zero then this fixes the small $k$ dependence
of the real part and we can use the Wronskian to infer how the imaginary
part depends upon $k$ and $a$,
\begin{equation}
{\rm Re}(u) \longrightarrow \frac{\#}{k^{\frac{D-1}2}} \qquad , \qquad 
{\rm Im}(u) \longrightarrow \frac1{2\# (D-1) H}
\frac{k^{\frac{D-1}2}}{a^{D-1}} \; .
\end{equation}
In this same limit the measure factor times retarded Green's function is 
the Fourier transform of the combination,
\begin{equation}
a^{\prime D-1} i \Bigl[u(t,k) u^*(t',k) - u^*(t,k) u(t',k)\Bigr] 
\longrightarrow \frac1{(D-1) H} \Bigl[1 - \Bigl(\frac{a'}{a}\Bigr)^{D-1} 
\Bigr] \; .
\end{equation}

In contrast, the leading small $k$ behavior of the conformally coupled 
mode functions is less singular than $u(t,k)$ by a factor of 
$(k/Ha)^{\frac{D}2-1}$. This precludes getting an infrared logarithm from 
the VEV of a pair of free fields. It also shifts scale factors from the
real part of $v$ to the imaginary part, 
\begin{equation}
{\rm Re}(v) \longrightarrow \frac{\#}{k^{\frac12} a^{\frac{D}2-1}} \qquad , 
\qquad {\rm Im}(v) \longrightarrow \frac1{2 \# H} \frac{k^{\frac12}}{
a^{\frac{D}2}} \; .
\end{equation}
Hence the measure factor times retarded Green's function goes to the Fourier 
transform of,
\begin{equation}
a^{\prime D-1} i \Bigl[ v(t,k) v^*(t',k) - v^*(t,k) v(t',k) \Bigr] 
\longrightarrow \frac1{H} \Bigl[ \Bigl(\frac{a'}{a}\Bigr)^{\frac{D}2-1} -
\Bigl(\frac{a'}{a}\Bigr)^{\frac{D}2}\Bigr] \; .
\end{equation}
As we saw in the previous section, positive powers of $a'/a$ weight
temporal vertex integrations overwhelmly at their upper limits, which
prevents them from producing infrared logarithms. Similar considerations 
apply to all passive fields.

Derivatives also suppress infrared logarithms. For if a free active 
field is differentiated with respect to space then its mode sum contains 
an extra factor of $k$ and the logarithmic singularity evident in
(\ref{smallk}) is absent. A time derivative is even worse. In $D = 4$ it
gives rise to two extra factors of $k$,
\begin{equation}
\dot{u}(t,k) = \frac{\partial}{\partial t} \Biggl\{\frac{H}{\sqrt{2 k^3}} 
\Bigl[1 -\frac{ik}{H a}\Bigr] e^{\frac{ik}{Ha}} \Biggr\} = 
\frac{H}{\sqrt{2 k^3}} \Bigl[-\frac{k^2}{H a^2} \Bigr] e^{\frac{ik}{Ha}} \; .
\end{equation}
Of course it is possible that a differentiated field in the Lagrangian
will give rise to a differentiated vertex integration, rather than a 
differentiated free field, in the free field expansion. This also prevents 
the appearance of an extra infrared logarithm because the undifferentiated 
vertex integration is a function only of time at leading logarithm order. 
Hence a spatial derivative of it gives zero, whereas a time derivative would 
lower the number of infrared logarithms by one.

\begin{figure}
\centerline{\epsfig{file=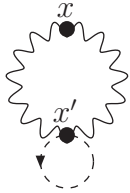}}
\caption{Two loop contribution to $\langle \Omega \vert F_{\mu\nu}(x) 
F_{\rho\sigma}(x)\vert \Omega \rangle$.}
\label{FF}
\end{figure}

Even though passive fields do not cause infrared logarithms, they can 
still transmit an infrared logarithm acquired through interaction with
an undifferentiated active field. Fig.~\ref{FF} depicts a two loop 
contribution to the VEV of $F_{\mu\nu}(x) F_{\rho\sigma}(x)$ in which 
an infrared logarithm from the coincident scalar loop at $x'$ is 
propagated through the virtual photon loop from $x'$ to $x$.

\begin{figure}
\centerline{\epsfig{file=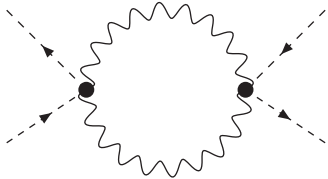}}
\caption{Effective $(\varphi^* \varphi)^2$ coupling in SQED.}
\label{quartic}
\end{figure}

\begin{figure}
\centerline{\epsfig{file=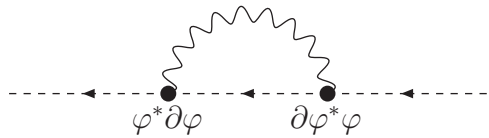}}
\caption{Effective $\varphi^* \varphi$ coupling in SQED.}
\label{phi2}
\end{figure}

Passive fields can also induce interactions between active fields. For 
example, the photon loop in Fig.~\ref{quartic} induces an effective 
$(\varphi^* \varphi)^2$ interaction,
The same comments apply to differentiated active fields. For example, 
Fig.~\ref{phi2} shows the differentiated scalar and the photon of the 3-point 
vertex (\ref{3pt}) inducing an interaction between undifferentiated
active fields.
This is part of the full 1PI 2-point function which has recently been 
computed at one loop order \cite{KW}.

\subsection{Reaching Leading Logarithm Order}

One can understand the relation between coupling constants and leading 
infrared logarithms directly from the Lagrangian. Consider a term in
the potential involving $N$ undifferentiated, active scalars, $V(\varphi) 
\sim c_N \varphi^N$. An elementary exercise in diagram topology reveals 
that it requires two such vertices to add $N\!-\!2$ loops to any diagram. 
Hence the loop counting parameter is $(C_N)^{\frac2{N-2}}$. Now consider 
the $2N$ extra fields associated with two $N$-point vertices. In 
the free field expansion some of these $2N$ fields would contribute to 
retarded Green's functions while others would remain as free fields. It 
requires two free fields to produce a retarded Green's function whose 
associated vertex integration can result in an infrared logarithm. An 
infrared logarithm can also come from the VEV of a pair of free fields. 
It follows that the $2N$ extra fields from two vertices can produce at 
most $N$ infrared logarithms. Hence the leading logarithm contributions 
to any VEV represent an expansion in powers of the parameter,
\begin{equation}
C_N \varphi^N \qquad \Longrightarrow \qquad \Bigl[C_N^2 \times \ln^N(a)
\Bigr]^{\frac1{N-2}} \; . \label{VtoLL}
\end{equation}
For $\lambda \varphi^3$ the series would be in powers of $\lambda^2 
\ln^3(a)$; for $\lambda \varphi^4$ we have already seen that the series
is in powers of $\lambda \ln^2(a)$; for $\lambda \varphi^5$ the series
would be in powers of $\lambda^2 \ln^5(a)$, and so on.

Now consider a model which consists of an active field $\varphi(x)$ and 
a passive field $\psi(x)$ that interact through a potential of the form, 
$K \psi^{\ell} (\partial \varphi)^m \varphi^n$. Because the interaction 
contains $N = \ell + m + n$ fields, the addition of two vertices to any 
diagram increases the number of loops by $N-2 = \ell + m + n - 2$. However, 
only the $2n$ active fields from these two vertices can contribute to 
infrared logarithms. Hence leading logarithm contributions to any VEV 
represent an expansion in the parameter,
\begin{equation}
K \psi^{\ell} (\partial \varphi)^m \varphi^n \qquad \Longrightarrow \qquad 
\Bigl[ K^2 \times \ln^n(a)\Bigr]^{\frac1{\ell+m+n-2}} \; . \label{genV}
\end{equation}
For example, the $\kappa h \partial h \partial h$ vertex of quantum
gravity has $\ell = 0$, $m=2$ and $n=1$, which produces a series in
powers of $\kappa^2 \ln(a)$. The same result follows for any of the
$\kappa^n h^n \partial h \partial h$ interactions of quantum gravity.

A model whose leading logarithm solution has already been obtained is 
Yukawa theory \cite{MW1},
\begin{eqnarray}
\lefteqn{\mathcal{L} = -\frac12 (1 + \delta Z) \partial_{\mu} \varphi 
\partial_{\nu} \varphi g^{\mu\nu} \sqrt{-g} - \frac{\delta \xi}2 \varphi^2 R 
\sqrt{-g} - \frac{\delta \lambda}{4!} \varphi^4 \sqrt{-g} } \nonumber \\
& & \hspace{.5cm} + (1 + \delta Z_2) i \overline{\psi} e^{\mu}_{~b} \gamma^b
\Bigl(\partial_{\mu} + \frac{i}2 A_{\mu c d} J^{cd} \Bigr) \psi \sqrt{-g}
- (f + \delta f) \varphi \overline{\psi} \psi \sqrt{-g} \; . \qquad
\end{eqnarray}
The scalar $\varphi$ is active whereas the massless fermion $\psi$ is
passive. The basic interaction vertex is $-f \varphi \overline{\psi} \psi 
\sqrt{-g}$ and comparison with (\ref{genV}) shows that leading logarithm
contributions represent an expansion in powers of $f^2 \ln(a)$. Contrast
this with a $-f \varphi^3 \sqrt{-g}$ vertex which would produce an expansion 
in powers of $f^2 \ln^3(a)$. So we see that {\it trading in an active field 
for a passive field, or for a differentiated active field, reduces the number
of infrared logarithms per coupling constant. However, there will still be
infrared logarithms as long as the vertex contains any undifferentiated 
active fields.}

\subsection{The Role of the Ultraviolet}

It is instructive to examine the behavior of counterterms in Yukawa theory.
Recall that they completely drop out at leading logarithm order when only 
undifferentiated active fields are present. The various counterterms of
Yukawa theory behave as follows,
\begin{equation}
\delta Z \sim f^2 \quad , \quad \delta Z_2 \sim f^2 \quad , \quad
\delta f \sim f^3 \quad , \quad \delta \xi \sim f^2 \quad , \quad
\delta \lambda \sim f^4 \; .
\end{equation}
Field strength renormalization is irrelevant because the incorporation of 
either a $\delta Z$ or a $\delta Z_2$ vertex would add a factor of $f^2$ 
without any undifferentiated active fields. The 3-point counterterm $\delta f$
is also subleading because it carries three extra factors of $f$ with only 
one undifferentiated active field. However, the one loop contribution to the 
conformal counterterm $\delta \xi$ adds a factor of $f^2$ with two 
undifferentiated active fields. These two fields could produce the extra
$\ln(a)$ needed to remain at leading logarithm order. Similarly, the one 
loop contribution to the 4-point counterterm $\delta \lambda$ can also give 
leading logarithm corrections. Note that in neither case do we need to worry 
about higher loop counterterms because these would bring more factors of 
$f^2$ with no additional active fields. So we see that {\it leading logarithm
order in Yukawa theory requires one loop conformal and 4-point counterterms,
but no other counterterms, and no renormalization at all beyond one loop.}

Passive fields engender ultraviolet divergences, even at leading logarithm
order, precisely because they do not produce infrared logarithms. What they
give instead is factors of order one which multiply the infrared logarithms
from active fields. It is important to understand that these factors of
order one derive from the full range of the free field mode sum and from
the full structure of the mode functions. Therefore, it is not possible to
stochastically simplify Heisenberg operator equations which contain passive
fields. The same considerations apply to differentiated active fields.

\subsection{Purging the Passive Fields}

It might seem as if there is no tractable formalism that describes the 
leading logarithm limit of models which contain passive fields. This is 
false. By integrating out the passive fields one obtains effective field 
equations which involve only active fields. Although these equations are
hideously nonlocal, we will explain how they are equivalent to simple,
local equations at leading logarithm order.

Effective field equations are notoriously difficult on account of the
field-dependent, inverse differential operators they involve. For example,
consider integrating out a conformal scalar $\psi(x)$ whose Lagrangian is,
\begin{equation}
{\cal L} = -\frac12 \partial_{\mu} \psi \partial_{\nu} \psi g^{\mu\nu} 
\sqrt{-g} -\frac18 \Bigl(\frac{D-2}{D-1}\Bigr) \psi^2 R \sqrt{-g}
-\frac{\lambda}4 \varphi^2 \psi^2 \sqrt{-g} \; .
\end{equation}
The resulting contribution to the $\varphi(x)$ equation of motion takes the
form,
\begin{equation}
-\lambda \varphi(x) \Bigl\langle x \Bigl\vert \frac{i}{\partial_{\mu}
(\sqrt{-g} g^{\mu\nu} \partial_{\nu}) - \frac14 (\frac{D-2}{D-1}) R
\sqrt{-g} - \frac{\lambda}{2} \varphi^2 \sqrt{-g} } \Bigr\vert x \Bigr\rangle
\; , \label{V'}
\end{equation}
where the quantity at the right is the coincidence limit of the $\psi$
propagator in an arbitrary $\varphi(x)$ background.\footnote{It should
properly be the coincidence limit of the $++$ propagator of the 
Schwinger-Keldysh formalism \cite{RJ} so that it depends only upon fields
$\varphi(x')$ in the past light-cone of $x^{\mu}$. Giving this a proper 
explication would require a substantial digression which we shall forgo.}
We shall probably never know this Green's function for arbitrary 
$\varphi(x)$. It can be expanded in terms of the conformal propagator 
$i\Delta_{\rm cf}(x;x')$ for $\varphi = 0$,
\begin{eqnarray}
\lefteqn{ \Bigl\langle x \Bigl\vert \frac{i}{\partial_{\mu} (\sqrt{-g} 
g^{\mu\nu} \partial_{\nu}) - \frac14 (\frac{D-2}{D-1}) R \sqrt{-g} - 
\frac{\lambda}{2} \varphi^2 \sqrt{-g} } \Bigr\vert x \Bigr\rangle  = 
i \Delta_{\rm cf}(x;x) } \nonumber \\
& & \hspace{-.5cm} - \frac{i\lambda}2 \int d^Dx' \sqrt{g(x')} \, \varphi^2(x') 
\Bigl[i\Delta_{\rm cf}(x;x')\Bigr]^2 + \Bigl(-\frac{i \lambda}2\Bigr)^2 
\int d^Dx' \sqrt{-g(x')} \varphi^2(x') \nonumber \\
& & \hspace{.5cm} \times \int d^Dx'' \, \sqrt{-g(x'')} \, \varphi^2(x'')
i\Delta_{\rm cf}(x;x') i\Delta_{\rm cf}(x';x'') i\Delta_{\rm cf}(x'';x) 
+ \ldots \qquad \label{expand}
\end{eqnarray}
We do know $i\Delta_{\rm cf}(x;x')$ --- in conformal coordinates it is just
$(a a')^{1-\frac{D}2}$ times the massless, flat space propagator --- but
there is no way of doing the integrals and summing the series for arbitrary
$\varphi(x')$.

That we can obtain a tractable formalism at leading logarithm order derives
from two facts:
\begin{enumerate}
\item{To reach leading logarithm order, every active field must contribute
to an infrared logarithm; and}
\item{Integrating over passive field Green's functions cannot produce infrared
logarithms.}
\end{enumerate}
The first fact means we can ignore the spatial dependence of all the 
$\varphi$'s in expansion (\ref{expand}). Of course their temporal 
dependence matters because this is the ultimate source of infrared logarithms.
However, integrating infrared logarithms against a passive field Green's 
function does not change the leading infrared logarithm. For example,
compare the result of having infrared logarithms inside such an integral
with the result of placing them outside,
\begin{eqnarray}
\lefteqn{\int_0^t dt' a^{\prime D-1} \int d^{D-1}x' G_{\rm cf}(x;x') \ln^N(a')}
\nonumber \\
& & \hspace{-.5cm} = \frac1{H} \int_0^t dt' \Biggl[ \Bigl(\frac{a'}{a}\Bigr)^{
\frac{D}2-1} \!\!\!\!\!- \Bigl(\frac{a'}{a}\Bigr)^{\frac{D}2} \Biggr] \ln^N(a')
= \frac{4 \ln^N(a)}{D (D-2) H^2} \Biggl\{1 + O\Bigl(\frac1{\ln(a)}\Bigr) 
\Biggr\} , \quad \\
\lefteqn{\ln^N(a) \int_0^t dt' a^{\prime D-1} \int d^{D-1}x' G_{\rm cf}(x;x')}
\nonumber \\
& & \hspace{-.5cm} = \frac{\ln^N(a)}{H} \int_0^t dt' \Biggl[\Bigl(\frac{a'}{a} 
\Bigr)^{\frac{D}2-1} - \Bigl(\frac{a'}{a}\Bigr)^{\frac{D}2} \Biggr] = 
\frac{4 \ln^N(a)}{D (D-2) H^2} \Biggl\{1 + O\Bigl(a^{1-\frac{D}2}\Bigr) 
\Biggr\} . \;
\end{eqnarray}
To leading logarithm order there is no difference! 

>From the preceding discussion we see that the leading logarithms are not
changed by moving all the $\varphi$'s of expansion (\ref{expand}) from 
inside the various integrals to outside, 
\begin{eqnarray}
\lefteqn{ \Bigl\langle x \Bigl\vert \frac{i}{\partial_{\mu} (\sqrt{-g} 
g^{\mu\nu} \partial_{\nu}) - \frac14 (\frac{D-2}{D-1}) R \sqrt{-g} - 
\frac{\lambda}{2} \varphi^2 \sqrt{-g} } \Bigr\vert x \Bigr\rangle  
\longrightarrow i \Delta_{\rm cf}(x;x) } \nonumber \\
& & \hspace{-.5cm} - \frac{i\lambda}2 \varphi^2(x) \int d^Dx' \sqrt{g(x')} 
\Bigl[i\Delta_{\rm cf}(x;x')\Bigr]^2 + \Bigl(-\frac{i \lambda}2 \varphi^2(x)
\Bigr)^2 \int d^Dx' \sqrt{-g(x')} \nonumber \\
& & \hspace{.5cm} \times \int d^Dx'' \, \sqrt{-g(x'')} i\Delta_{\rm cf}(x;x') 
i\Delta_{\rm cf}(x';x'') i\Delta_{\rm cf}(x'';x) + \ldots \; .
\end{eqnarray}
This last expression can be recognized as the coincidence limit of the
propagator of a conformally coupled scalar with mass $m^2 = \frac{\lambda}2
\varphi^2(x)$. A compact result for this can be given in terms of the parameter
$\nu \equiv \frac12 \sqrt{1 - 2 \lambda \varphi^2(x)/H^2}$ \cite{PCDJR,JSDRC},
\begin{eqnarray}
\lefteqn{ \Bigl\langle x \Bigl\vert \frac{i}{\partial_{\mu} (\sqrt{-g} 
g^{\mu\nu} \partial_{\nu}) - \frac14 (\frac{D-2}{D-1}) R \sqrt{-g} - 
\frac{\lambda}{2} \varphi^2 \sqrt{-g} } \Bigr\vert x \Bigr\rangle }\nonumber \\
& & \hspace{-.3cm} \longrightarrow \frac{H^{D-2}}{(4\pi)^{\frac{D}2}} 
\frac{\Gamma(\frac{D-1}2 + \nu) \Gamma(\frac{D-1}2 - \nu)}{\Gamma(\frac{D}2)} 
\mbox{}_2F_1\Bigl(\frac{D-1}2 + \nu, \frac{D-1}2 - \nu; \frac{D}2 ; 1\Bigr) 
\; , \qquad \\
& & = \frac{H^{D-2}}{(4\pi)^{\frac{D}2}} \frac{\Gamma(1 - \frac{D}2)
\Gamma(\frac{D-1}2 + \nu) \Gamma(\frac{D-1}2 - \nu)}{\Gamma(\frac12 + \nu)
\Gamma(\frac12 - \nu)} \; . \label{coinc}
\end{eqnarray}

It remains to substitute (\ref{coinc}) in (\ref{V'}) and recognize the result
as minus the derivative of the unrenormalized, one loop effective potential,
\begin{equation}
-V'_{\rm eff}(\varphi) = -\lambda \varphi(x) \times
\frac{H^{D-2}}{(4\pi)^{\frac{D}2}} \frac{\Gamma(1 - \frac{D}2)
\Gamma(\frac{D-1}2 + \nu) \Gamma(\frac{D-1}2 - \nu)}{\Gamma(\frac12 + \nu)
\Gamma(\frac12 - \nu)} \; .
\end{equation}
The factor of $\Gamma(1-\frac{D}2)$ represents a one loop, ultraviolet 
divergence which can be absorbed by renormalizing the quartic conformal 
and quartic self-couplings. The model that results is of the scalar
potential type (\ref{Starform}) already solved by Starobinski\u{\i}.

A little reflection reveals that the technique we have just sketched must
always work. Of course we can integrate out any subset of fields. The 
nonlocality engendered by doing this to passive fields will necessarily
be restricted to passive field Green's functions. By definition of the 
field being passive, such Green's functions cannot produce infrared 
logarithms. Hence it is always valid to extract undifferentiated active 
fields from inside the nonlocal effective field equations. The result must 
always give minus the derivative of the unrenormalized effective potential.

A final point should be noted concerning expectation values of operators,
in the original theory, which involve passive fields. In computing the 
leading logarithm result for such an expectation value one must functionally
integrate out the passive fields. This will produce a potentially divergent
expression involving only active fields. The stochastic expectation value
of this expression can then be computed using the effective potential.

\section{Effective Potential of SQED}

The point of this paper is to obtain a leading logarithm solution for
SQED and that is largely accomplished in this section. We begin by applying 
the procedure of section 3.2 to show that SQED gives a series in powers of
$e^2 \ln(a)$ at leading logarithm order. We also work out which counterterms 
make leading log contributions. We then integrate out the photon field and 
renormalize the scalar effective potential. The section closes with large 
field and small field expansions of the effective potential.

We repeat the SQED Lagrangian (\ref{L}) from the Introduction,
\begin{eqnarray}
{\cal L} &\! = \!&
- (1 + \delta Z_2) \, (\partial_{\mu} - ieA_{\mu}) \varphi^* \,
(\partial_{\nu} + ieA_{\nu}) \varphi \; g^{\mu\nu} \sqrt{-g}
- \; \delta\xi \, \varphi^* \varphi \; R \, \sqrt{-g} \;
\nonumber \\
& \mbox{} &
- \, \frac14 (1 + \delta Z_3) \, F_{\mu\nu} \, F_{\rho\sigma} \,
g^{\mu\rho} \, g^{\nu\sigma} \sqrt{-g} \;
- \, \frac{\delta\lambda}{4} \, (\varphi^* \varphi)^2 \, \sqrt{-g}
\;\; , 
\end{eqnarray}
The complex scalar $\varphi(x)$ is active whereas the photon $A_{\mu}(x)$ is
passive. The primitive 3-point interaction involves a passive field, a
differentiated active field and an undifferentiated active field, so we
have the case of (\ref{VtoLL}) with $\ell = m = n = 1$,
\begin{equation}
ie A_{\mu} \Bigl[ \varphi^* \partial_{\nu} \varphi - \partial_{\nu} \varphi^*
\varphi\Bigr] g^{\mu\nu} \sqrt{-g} \qquad \Longrightarrow \qquad e^2 \ln(a)\; .
\label{3pt}
\end{equation}
The primitive 4-point interaction represents $\ell = n = 2$ with $m=0$, so
the result is the same,
\begin{equation}
-e^2 A_{\mu} A_{\nu} \varphi^* \varphi g^{\mu\nu} \sqrt{-g} \qquad 
\Longrightarrow \qquad e^2 \ln(a) \; .
\end{equation}

The various counterterms have the following dependences upon $e^2$,
\begin{equation}
\delta Z_2 \sim e^2 \quad , \quad \delta Z_3 \sim e^2 \quad , \quad 
\delta \xi \sim e^2 \quad , \quad \delta \lambda \sim e^4 \; .
\end{equation}
Hence field strength renormalization cannot contribute at leading logarithm
order because the $\delta Z_2$ and $\delta Z_3$ interactions add a factor 
of $e^2$ with no undifferentiated active fields. On the other hand, the 
conformal and quartic counterterms correspond to $\ell = m = 0$, with $n = 
2$ and $n=4$, respectively. We therefore conclude that {\it leading logarithm
SQED requires one loop conformal and 4-point counterterms, but no other
counterterms and no renormalization at all beyond one loop.}

It is now time to integrate out the vector potential. Ever since the classic
work of Coleman and Weinberg \cite{SRCEW}, it has been realized that this is
is greatly facilitated in Lorentz gauge,
\begin{equation}
\partial_{\mu} \Bigl( \sqrt{-g} g^{\mu\nu} A_{\nu} \Bigr) = 0 \; . 
\label{Lorentz}
\end{equation}
The simplifications turn out to be even greater in de Sitter background so 
we will follow the usual practice, although other gauges can of course be 
employed \cite{Jackiw}. Dropping the field strength renormalizations, 
partially integrating, using the gauge condition and expanding the Lagrangian 
in powers of the vector potential gives,
\begin{eqnarray}
{\cal L} & \longrightarrow & {\cal L}_0 + {\cal L}_1 + {\cal L}_2 \; , \\
{\cal L}_0 & = & -\partial_{\mu} \varphi^* \partial_{\nu} \varphi g^{\mu\nu}
\sqrt{-g} - \delta \xi \varphi^* \varphi R \sqrt{-g} -\frac{\delta \lambda}4
(\varphi^* \varphi)^2 \sqrt{-g} \; , \qquad \label{L0} \\
{\cal L}_1 & = & - \Bigl[\partial_{\mu} \varphi^* \varphi - \varphi^* 
\partial_{\mu} \varphi\Bigr] i e A_{\nu} g^{\mu\nu} \sqrt{-g} \equiv J^{\nu}
A_{\nu} \; \label{L1} , \\
{\cal L}_2 & = & \frac12 A_{\mu} \Bigl[\square^{\mu\nu} - R^{\mu\nu} - 2 e^2
\varphi^* \varphi g^{\mu\nu} \Bigr] A_{\nu} \sqrt{-g} \; . \label{L2}
\end{eqnarray}
Here $\square^{\mu\nu}$ is the vector d'Alembertian defined by $\square^{\mu
\nu} A_{\nu} = A^{\mu ; \rho}_{~~~\rho}$. The effective action is determined
by the equation,
\begin{equation}
e^{i \Gamma[\varphi^*,\varphi]} \equiv \Fint [dA_{\mu}] 
\delta\Bigl[\partial_{\mu}(\sqrt{-g} g^{\mu\nu} A_{\nu}) \Bigr] 
e^{i S[\varphi^*,\varphi,A]} \; .
\end{equation}
Because ${\cal L}$ is quadratic in the vector potential we can obtain the
following explicit expression for $\Gamma[\varphi^*,\varphi]$,
\begin{eqnarray}
\lefteqn{\Gamma[\varphi^*,\varphi] = S_0[\varphi^*,\varphi] + \frac{i}2 
\ln\Biggl\{ \det\Biggl[ \sqrt{-g} \Bigl(\square^{\mu\nu} - R^{\mu\nu} - 2 e^2
\varphi^* \varphi g^{\mu\nu}\Bigr)\Biggr] \Biggr\} } \nonumber \\
& & \frac{i}2 \int d^Dx J^{\mu}(x) \int d^Dy \Bigl\langle x\Bigl\vert 
\frac{i}{\sqrt{-g} [\square^{\mu\nu} - R^{\mu\nu} - 2 e^2 \varphi^* \varphi 
g^{\mu\nu}]} \Bigr\vert y \Bigr\rangle J^{\nu}(y) \; . \qquad \label{Gamma}
\end{eqnarray}

Expression (\ref{Gamma}) consists of three terms: the purely scalar parts
of the bare action, a determinant factor, and the ``current-current'' term
from completing the square in the exponential of the functional integral. 
One of the great things about Lorentz gauge is that this third term drops 
out at leading logarithm order. Recall from the previous section that 
leading logarithm contributions are unchanged by moving undifferentiated
scalars around to different sides of inverse differential operators. But
then a partial integration gives a gradient of Lorentz gauge propagator,
which vanishes. For example, consider the leading logarithm contribution 
from the second term in $J^{\mu}(x)/ie$,
\begin{eqnarray}
\lefteqn{\int \!\! d^Dx \, \varphi^*(x) \partial_{\rho} \varphi(x) \sqrt{-g(x)}
g^{\rho\mu}(x) \Bigl\langle x\Bigl\vert \frac{i}{\sqrt{-g} [\square^{\mu\nu} 
\!-\! R^{\mu\nu} \!-\! 2 e^2 \varphi^* \varphi g^{\mu\nu}]} \Bigr\vert y 
\Bigr\rangle \Biggl\vert_{\rm lead\ log} } \nonumber \\
& & \hspace{-.7cm} = \varphi^*(y) \!\! \int \!\! d^Dx \, \partial_{\rho} 
\varphi(x) \sqrt{-g(x)} g^{\rho\mu}(x) \Bigl\langle x\Bigl\vert 
\frac{i}{\sqrt{-g} [\square^{\mu\nu} \!-\! R^{\mu\nu} \!-\! 2 e^2 \varphi^* 
\varphi g^{\mu\nu}]} \Bigr\vert y \Bigr\rangle , \qquad \label{step1} \\
& & \hspace{-.7cm} = -\varphi^* \!\! \int \!\! d^Dx \, \varphi \partial_{\rho} 
\Biggl[ \sqrt{-g} g^{\rho\mu} \Bigl\langle x\Bigl\vert \frac{i}{\sqrt{-g} 
[\square^{\mu\nu} \!-\! R^{\mu\nu} \!-\! 2 e^2 \varphi^* \varphi g^{\mu\nu}]} 
\Bigr\vert y \Bigr\rangle \Biggr] = 0 \; . \qquad \label{step2}
\end{eqnarray}
The same cancellation also occurs for the flat space effective potential
\cite{SRCEW}.

The determinant term in (\ref{Gamma}) makes a leading logarithm contribution
to the effective field equations, but only in the form of the coincidence 
limit of a massive photon propagator, with $m^2 \equiv 2 e^2 \varphi^* \varphi$
treated as if it were constant. The massive photon propagator and its 
coincidence limit have recently been worked out in de Sitter background 
\cite{TW4} and the result is,
\begin{eqnarray}
\lefteqn{ \frac{\delta \Gamma[\varphi^*,\varphi]}{\delta \varphi^*(x)}
\longrightarrow \frac{\delta S_0[\varphi^*,\varphi]}{\delta \varphi^*(x)} }
\nonumber \\
& & -e^2 \varphi(x) \sqrt{-g} g^{\mu\nu} \Bigl\langle x \Bigl\vert
\frac{i}{\sqrt{-g} [\square^{\mu\nu} - R^{\mu\nu} - 2 e^2 \varphi^* \varphi
g^{\mu\nu}]} \Bigr\vert x \Bigr\rangle \; , \\
& & \hspace{-.5cm} \longrightarrow \partial_{\mu} \Bigl(\sqrt{-g} g^{\mu\nu} 
\partial_{\nu} \varphi\Bigr) - \delta \xi \varphi R \sqrt{-g} - 
\frac{\delta \lambda}2 \varphi^* \varphi^2 \sqrt{-g} -e^2 \varphi \sqrt{-g} 
g^{\mu\nu} \nonumber \\
& & \times g_{\mu\nu}
\Bigl(\frac{D\!-\!1}2\Bigr) \frac{H^2}{m^2} \frac{H^{D-2}}{(4\pi)^{\frac{D}2}}
\Biggl\{\! \frac{\Gamma(D\!-\!1)}{\Gamma(\frac{D}2\! +\! 1)} - \frac{\Gamma(-
\frac{D}2) \Gamma(\frac{D+1}2 \!+\! \nu) \Gamma(\frac{D+1}2 \!-\! \nu)}{
\Gamma(\frac12 + \nu) \Gamma(\frac12 - \nu)} \!\Biggr\} . \qquad \label{Veqn}
\end{eqnarray}
The parameter $\nu$ has the following definition,
\begin{equation}
\nu \equiv \sqrt{\Bigl(\frac{D-3}2\Bigr)^2 - \frac{m^2}{H^2}} \longrightarrow
\sqrt{\Bigl(\frac{D-3}2\Bigr)^2 - \frac{2 e^2 \varphi^* \varphi}{H^2}} \; .
\label{nudef}
\end{equation}

We can recognize the derivative of the unrenormalized effective potential 
from expression (\ref{Veff}),
\begin{eqnarray}
\lefteqn{V'_{\rm eff}(\varphi^* \varphi) = \delta \xi D (D\!-\!1) H^2 + 
\frac{\delta \lambda}2 \varphi^* \varphi + \frac{e^2}2 D (D\!-\!1) 
\frac{H^{D-2}}{(4\pi)^{\frac{D}2}} } \nonumber \\
& & \hspace{3cm} \times \frac{H^2}{m^2} \Biggl\{\frac{\Gamma(D\!-\!1)}{
\Gamma(\frac{D}2 \!+\! 1)} - \Gamma\Bigl(-\frac{D}2\Bigr) \frac{\Gamma(
\frac{D+1}2 \!+\! \nu) \Gamma(\frac{D+1}2 \!-\! \nu)}{\Gamma(\frac12 + \nu) 
\Gamma(\frac12 - \nu)} \Biggr\} . \qquad \label{Vprime}
\end{eqnarray}
Renormalization is accomplished by first setting $D = 4 - \epsilon$ and 
expanding the final term,
\begin{eqnarray}
\lefteqn{ \frac{H^2}{m^2} \Biggl\{\frac{\Gamma(3\!-\!\epsilon)}{\Gamma(3\!-\!
\frac{\epsilon}2)} - \Gamma\Bigl(-2 \!+\! \frac{\epsilon}2\Bigr) 
\frac{\Gamma(\frac52 \!+\! \nu \!-\! \frac{\epsilon}2) \Gamma(\frac52 \!-\! 
\nu \!-\! \frac{\epsilon}2)}{\Gamma(\frac12 + \nu) \Gamma(\frac12 - \nu)} 
\Biggr\} = -\Bigl(1 + \frac{e^2 \varphi^* \varphi}{H^2} \Bigr) 
\frac{2}{\epsilon} } \nonumber \\
& & \hspace{1.5cm} + \frac12 + \Bigl(1 \!+\! \frac{e^2 \varphi^* \varphi}{H^2} 
\Bigr) \Biggl[ \psi\Bigl(\frac32 \!+\! \nu\Bigr) + \psi\Bigl(\frac32 \!-\! 
\nu\Bigr) -\frac32 + \gamma\Biggr] + O(\epsilon) \; . \qquad \label{gamexp}
\end{eqnarray}
The digamma function $\psi(z)$ has the following definition and expansions
for small $z$ and large $z$ \cite{GR},
\begin{eqnarray}
\psi(z) & \equiv & \frac{d}{dz} \ln\Bigl(\Gamma(z)\Bigr) \; , \\
& = & -\gamma + \sum_{n=2}^{\infty} (-1)^n \zeta(n) (z-1)^{n-1} \; , 
\label{smallz} \\
& = & \ln(z) - \frac1{2 z} - \frac1{12 z^2} + \frac1{120 z^4} 
+ O\Bigl(\frac1{z^6}\Bigr) \; . \label{largez}
\end{eqnarray}

We make the following choices for the two relevant counterterms,
\begin{eqnarray}
\delta \xi & = & \frac{e^2 H^{D-4}}{(4 \pi)^{\frac{D}2}} \Biggl\{
\frac1{4-D} + \frac{\gamma}{2} + O(D-4)\Biggr\} \; , \label{delxi} \\
\delta \lambda & = & \frac{(D-1) D e^4 H^{D-4}}{(4 \pi)^{\frac{D}2}} 
\Biggl\{ \frac2{4 -D} + \gamma - \frac32 + O(D-4)\Biggr\} \; . \label{dellam}
\end{eqnarray}
Substituting (\ref{gamexp}) and (\ref{delxi}-\ref{dellam}) in (\ref{Vprime})
and taking the limit $\epsilon \longrightarrow 0$ gives,
\begin{eqnarray}
\lefteqn{V'_{\rm eff}(\varphi^* \varphi) = \frac{e^2}{H^2} \times 
\frac{3 H^4}{8 \pi^2} \Biggl\{-1 \!+\! 2 \gamma + \Bigl(-3 \!+\! 2 \gamma\Bigr)
\frac{e^2 \varphi^* \varphi}{H^2} } \nonumber \\
& & \hspace{5cm} + \Bigl(1 \!+\! \frac{e^2 \varphi^* \varphi}{H^2}\Bigr)
\Biggl[\psi\Bigl(\frac32 \!+\! \nu\Bigr) + \psi\Bigl(\frac32 \!-\! \nu\Bigr)
\Biggr] \Biggr\} . \qquad \label{Vpren}
\end{eqnarray}
Of course the divergent parts of $\delta \xi$ and $\delta \lambda$ are fixed. 
Our choices for the finite parts are motivated to make the $(\varphi^* 
\varphi)^0$ and $(\varphi^* \varphi)^1$ terms in $V'_{\rm eff}(\varphi
\varphi^*)$ vanish, which keeps the scalar light as long as possible. 
Interestingly, the same choice for $\delta \xi$ cancels the leading infrared 
logarithm in the two loop expectation value of $\varphi^*(x) \varphi(x)$ 
\cite{PTsW}, and also keeps the one loop scalar mode functions from receiving 
any significant late time correction \cite{KW2}.

It remains to work out the effective potential and expand it for large
and small field strengths. From (\ref{Vpren}) and the definition (\ref{nudef}) 
of $\nu$ we see that the result depends upon the combination,
\begin{equation}
z \equiv \frac{e^2 \varphi^* \varphi}{H^2} \; .
\end{equation}
Integrating (\ref{Vpren}) gives,
\begin{eqnarray}
\lefteqn{V_{\rm eff} = \frac{3 H^4}{8 \pi^2} \Biggl\{ (-1 + 2 \gamma) z
+ (-\frac32 + \gamma) z^2 } \nonumber \\
& & \hspace{1.5cm} + \int_0^z \!\!\! dx \, (1 \!+\! x) \Biggl[ \psi\Bigl(
\frac32 \!+\! \frac12 \sqrt{1 \!-\! 8 x}\Bigr) + \psi\Bigl(\frac32 \!-\!
\frac12 \sqrt{1 \!-\! 8x}\Bigr)\Biggr] \Biggr\} . \qquad \label{Veff}
\end{eqnarray}
An explicit power series expansion can be obtained for $V_{\rm eff}$ in
terms of the parameter,
\begin{equation}
\Delta z \equiv \frac12 - \frac12 \sqrt{1 - 8 z} = 2 z + 4 z^2 + 16 z^3 
+ O(z^4) \; .
\end{equation}
Substituting (\ref{smallz}) in (\ref{Veff}) and performing the integral
gives,
\begin{eqnarray}
\lefteqn{V_{\rm eff} = \frac{3 H^4}{8 \pi^2} \Biggl\{ \frac12 \ln(1 - 
\Delta z) + \frac12 \Delta z + \frac14 {\Delta z}^2 + \frac7{12}
{\Delta z}^3 - \frac38 {\Delta z}^4 } \nonumber \\
& & + \sum_{m=1}^{\infty} \zeta(2m+1) \Biggl[-\frac{\Delta z^{2m+1}}{2m+1}
+ \frac{\frac32 {\Delta z}^{2m+2}}{2m+2} + \frac{\frac32 {\Delta z}^{2m+3}}{
2m+3} - \frac{\Delta z^{2m+4}}{2m+4} \Biggr] \Biggr\} , \qquad \\
& & = \Bigl[\frac5{12} -\frac13 \zeta(3)\Bigr] \Delta z^3 - \Bigl[\frac12
- \frac38 \zeta(3)\Bigr] \Delta z^4 + O(\Delta z^5) \; . \label{asymp}
\end{eqnarray}
As already stated, our choices for the finite parts of $\delta \xi$ and 
$\delta \lambda$ cancel the order $z$ and $z^2$ terms in the small field
expansion.

\begin{figure}
\centerline{\epsfig{file=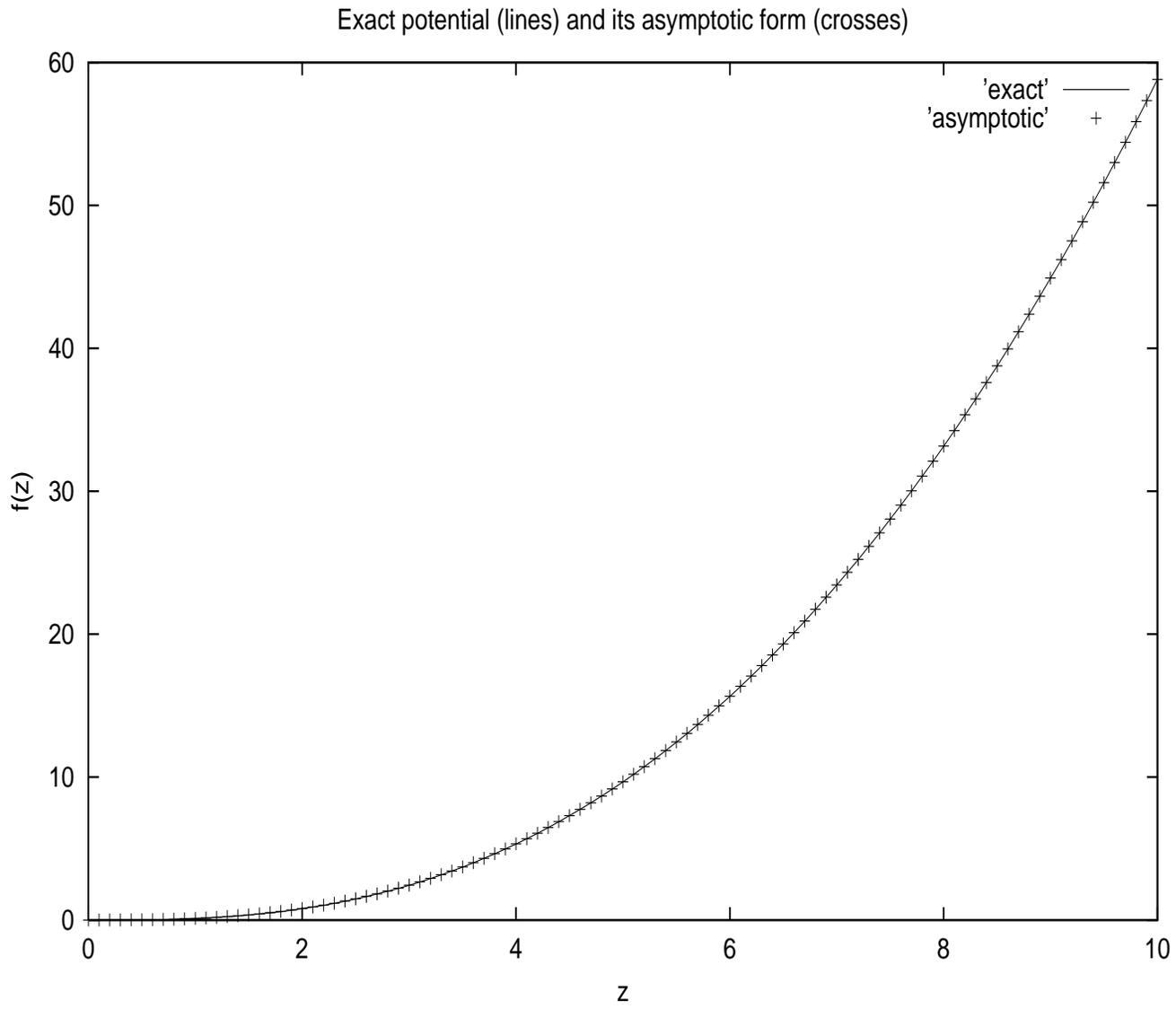,height=7in,width=6in,angle=180}}
\caption{$V_{\rm eff} = \frac{3 H^4}{8 \pi^2} f(\frac{e^2 \varphi^* \varphi}{
H^2})$ (solid line) and its asymptotic form (crosses).}
\label{full}
\end{figure}

\begin{figure}
\centerline{\epsfig{file=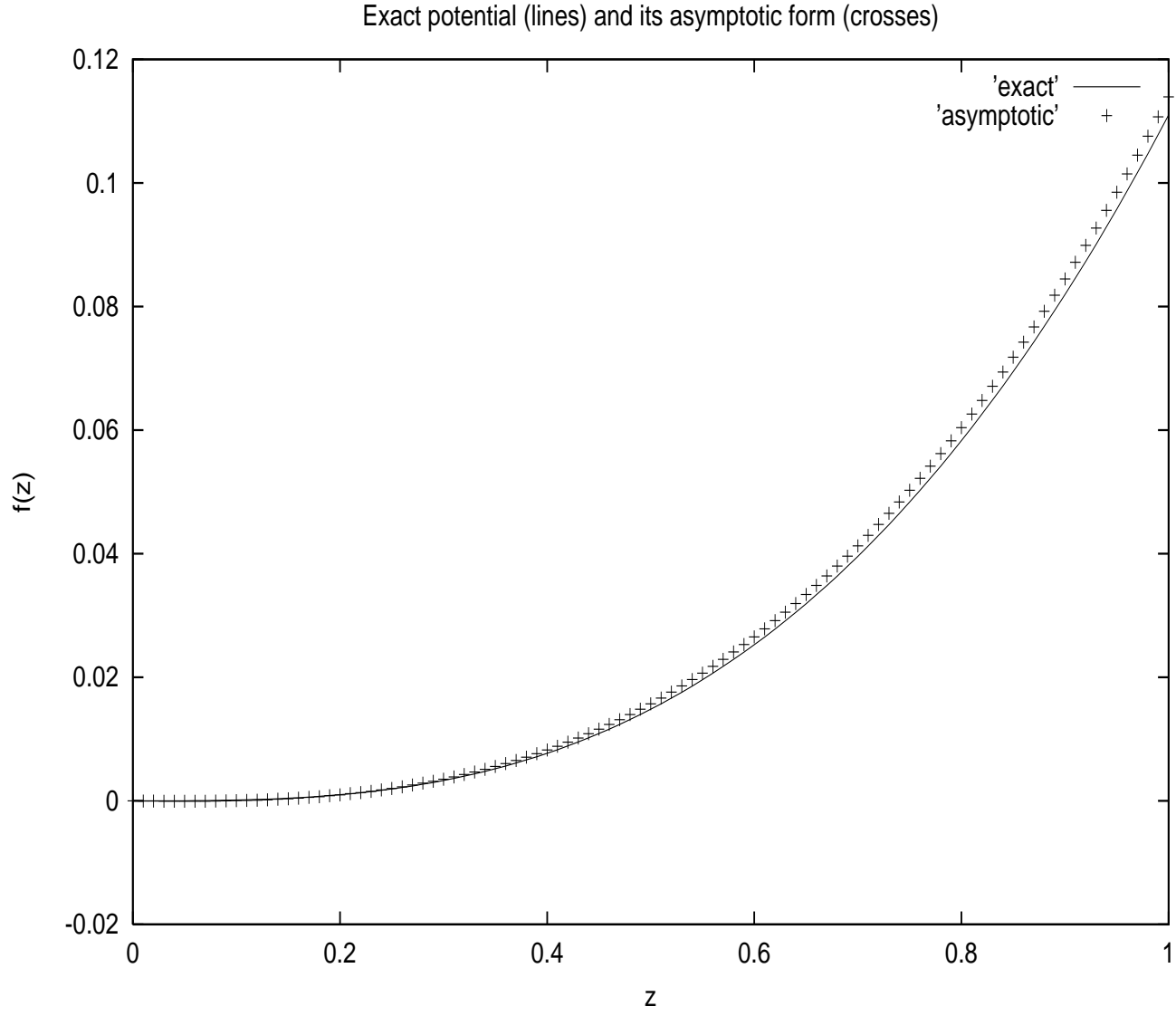,height=7in,width=6in,angle=180}}
\caption{Expanded small field behavior of $V_{\rm eff} = \frac{3 H^4}{8 \pi^2} 
f(\frac{e^2 \varphi^* \varphi}{H^2})$ (solid line) and its asymptotic form 
(crosses).}
\label{small}
\end{figure}

The large field expansion derives from substituting (\ref{largez}) in 
(\ref{Veff}),
\begin{eqnarray}
\lefteqn{V_{\rm eff} = \frac{3 H^4}{8 \pi^2} \Biggl\{ \frac12 z^2 \ln(z + 1)
+ \Bigl[-\frac74 + \frac12 \ln(2) + \gamma\Bigr] z^2 + z \ln(z + 1)}\nonumber\\
& & \hspace{3cm} + \Bigl[-\frac{13}6 + \ln(2) + 2 \gamma\Bigr] z + 
\frac{19}{60} \ln(z + 1) + O(1) \Biggr\} . \qquad \label{largephi}
\end{eqnarray}
Because $z$ grows for small $H$, as well as for large $\varphi^* \varphi$,
the $(\varphi^* \varphi)^2 \ln(\varphi^* \varphi)$ term in (\ref{largephi})
should also agree with the flat space result of Coleman and Weinberg
\cite{SRCEW,Gar}. From equation (4.5) of their paper we see that it does.
The full asymptotic expansion of (\ref{largephi}) up to order one also
gives a generally accurate approximation of the potential, even for $z < 1$,
as can be seen from Fig.~\ref{full} and Fig.~\ref{small}.

\section{SQED Stress Tensor}

The stress tensor of SQED is a composite operator which involves the
passive field $A_{\mu}$. One must therefore integrate $A_{\mu}$ out and
simplify the resulting functional of $\varphi^*$ and $\varphi$ before 
its expectation value can be computed stochastically. That is the task
of this section. As a bonus we obtain independent results for the two
gauge invariant operators which principally comprise $T_{\mu\nu}$: the 
field strength bilinear and the scalar kinetic bilinear. The section closes
with a discussion of the curious fact that the simplified stress tensor
is not quite $-g_{\mu\nu} V_{\rm eff}(\varphi^*\varphi)$.

Recall from the previous section that field strength renormalization
counterterms do not contribute at leading logarithm order whereas the
conformal and quartic counterterms do. The relevant part of the SQED
stress tensor is therefore,
\begin{eqnarray}
\lefteqn{T_{\mu\nu} = \Bigl[\delta^{\alpha}_{\mu} \delta^{\rho}_{\nu}
\!-\! \frac14 g_{\mu\nu} g^{\alpha\rho} \Bigr] g^{\beta\sigma} F_{\alpha\beta} 
F_{\rho \sigma} + \Bigl[\delta_{\mu}^{\rho} \delta_{\nu}^{\sigma} \!+\!
\delta_{\mu}^{\sigma} \delta_{\nu}^{\rho} \!-\! g_{\mu\nu} g^{\rho\sigma}
\Bigr] (D_{\rho} \varphi)^* D_{\sigma} \varphi } \nonumber \\
& & \hspace{-.5cm} + 2 \delta \xi \Bigl[\varphi^* \varphi \Bigl(R_{\mu\nu} 
\!-\! \frac12 g_{\mu\nu} R\Bigr) + g_{\mu\nu} (\varphi^* \varphi)^{; \rho}_{
~~ \rho} - (\varphi^* \varphi)_{;\mu\nu} \Bigr] - \frac{\delta \lambda}4 
(\varphi^* \varphi)^2 g_{\mu\nu} \; , \qquad \label{fullT}
\end{eqnarray}
where the covariant derivative is $D_{\mu} \varphi \equiv (\partial_{\mu} 
\varphi \!+\! i e A_{\mu} \varphi)$. In de Sitter background the Ricci
tensor is $R_{\mu\nu} = (D-1) H^2 g_{\mu\nu}$, and differentiated scalars 
with the same power of $e^2$ as undifferentiated scalars are guaranteed to 
be subleading logarithm. Hence we can simplify the counterterms to,
\begin{eqnarray}
\lefteqn{2 \delta \xi \Bigl[\varphi^* \varphi \Bigl(R_{\mu\nu} \!-\! \frac12 
g_{\mu\nu} R\Bigr) + g_{\mu\nu} (\varphi^* \varphi)^{; \rho}_{~~ \rho} - 
(\varphi^* \varphi)_{;\mu\nu} \Bigr] - \frac{\delta \lambda}4 (\varphi^* 
\varphi)^2 g_{\mu\nu} } \nonumber \\
& & \hspace{3cm} \longrightarrow - \Bigl[ (D \!-\! 1) (D \!-\! 2) \delta \xi
H^2 \varphi^* \varphi + \frac{\delta \lambda}4 (\varphi^* \varphi)^2\Bigr]
g_{\mu\nu} \; . \qquad \label{ctms}
\end{eqnarray}
It remains to integrate the vector potential out of the field strength
and scalar kinetic bilinears,
\begin{equation}
F_{\alpha\beta} F_{\rho\sigma} \qquad {\rm and} \qquad (D_{\rho} \varphi)^*
D_{\sigma} \varphi \; .
\end{equation}

Consider the general case of integrating the vector potential out of some 
operator ${\cal O}[\varphi^*,\varphi,A]$ to obtain a new operator
$\widetilde{\cal O}[\varphi^*,\varphi]$ depending only upon the scalar,
\begin{equation}
\Fint [dA_{\mu}] \delta\Bigl[\partial_{\mu}(\sqrt{-g} g^{\mu\nu} A_{\nu}) 
\Bigr] e^{i S[\varphi^*,\varphi,A]} \times {\cal O}[\varphi^*,\varphi,A]
= e^{i \Gamma[\varphi^*,\varphi]} \times \widetilde{\cal O}[\varphi^*,\varphi]
\; . \label{Oint}
\end{equation}
The functional integration is trivial because the Lagrangian of SQED is 
quadratic in the vector potential. After some partial integrations and 
applications of the Lorentz gauge condition (\ref{Lorentz}), it consists 
of a part ${\cal L}_0$ which depends only upon the scalar, a linear part of 
the form ${\cal L}_1 = J^{\nu} A_{\nu}$, and a quadratic part ${\cal L}_2 = 
\frac12 A_{\mu} {\cal D}^{\mu\nu} A_{\nu}$. We can read off the current 
$J^{\nu}$ from (\ref{L1}) and the differential operator ${\cal D}^{\mu\nu}$ 
from (\ref{L2}),
\begin{eqnarray}
J^{\nu} & \equiv & -ie \Bigl[\partial_{\mu} \varphi^* \varphi \!-\! \varphi^*
\partial_{\mu} \varphi\Bigr] g^{\mu\nu} \sqrt{-g} \; , \\
{\cal D}^{\mu\nu} & \equiv & \sqrt{-g} \Bigl[ \square^{\mu\nu} - R^{\mu\nu} 
- 2 e^2 \varphi^* \varphi g^{\mu\nu} \Bigr] \; . \label{scriptD}
\end{eqnarray}
One evaluates the functional integral (\ref{Oint}) by completing the square,
\begin{equation}
\frac12 A_{\mu} {\cal D}^{\mu\nu} A_{\nu} + A_{\nu} J^{\nu}
= \frac12 \Bigl[A_{\mu} + \frac1{{\cal D}^{\mu\rho}} J^{\rho} \Bigr]
{\cal D}^{\mu\nu} \Bigl[A_{\nu} + \frac1{{\cal D}^{\nu\sigma}} J^{\sigma} 
\Bigr] - \frac12 J^{\mu} \frac1{{\cal D}^{\mu\nu}} J^{\nu} \; .
\end{equation}
The operator $\widetilde{\cal O}[\varphi^*,\varphi]$ will therefore be the
sum of terms from ${\cal O}[\varphi^*,\varphi,A]$ in which all combinations
of the following replacements are made,
\begin{eqnarray}
A_{\mu}(x) & \longrightarrow & -\int d^Dx' \, \Bigl\langle x \Bigl\vert 
\frac1{{\cal D}^{\mu\nu}} \Bigr\vert x' \Bigr\rangle J^{\nu}(x') \; , 
\label{rep1} \\
A_{\mu}(x) A_{\nu}(x') & \longrightarrow & \Bigl\langle x \Bigl\vert
\frac{i}{{\cal D}^{\mu\nu}} \Bigr\vert x' \Bigr\rangle \; . \label{rep2}
\end{eqnarray} 
As we saw from expressions (\ref{step1}-\ref{step2}) of the previous section,
the Lorentz gauge condition means that there is never a leading logarithm 
contribution from replacement (\ref{rep1}). Hence we need only consider
replacement (\ref{rep2}).

At this stage we must digress to discuss the photon propagator. We shall 
never know the inverse of the differential operator (\ref{scriptD}) for
arbitrary $\varphi(x)$. However, leading logarithm results
only involve this operator evaluated for the special case where $2 e^2
\varphi^*(x) \varphi(x)$ is a constant we shall call $m^2$. That Green's
function we do know \cite{TW4}. It can be expressed in terms of the 
de Sitter invariant function of conformal coordinates $x^{\mu} = 
(\eta,\vec{x})$,
\begin{equation}
y(x;x') \equiv a a' H^2 \Bigl[\Vert \vec{x} - \vec{x}' \Vert^2 - 
\Bigl(\vert \eta - \eta'\vert - i \epsilon\Bigr)^2 \; .
\end{equation}
The massive, Lorentz gauge photon propagator takes the form \cite{TW4},
\begin{equation}
\Bigl\langle x \Bigl\vert \frac{i}{{\cal D}^{\mu\nu}} \Bigr\vert x' 
\Bigr\rangle = B(y) \frac{\partial^2 y}{\partial x^{\mu} \partial 
x^{\prime\nu}} + C(y) \frac{\partial y}{\partial x^{\mu}} \frac{\partial y}{
\partial x^{\prime \nu}} \; .
\end{equation}
The functions $B(y)$ and $C(y)$ can be expressed in terms of a single 
function $\gamma(y)$,
\begin{eqnarray}
B(y) & \equiv & \frac1{4 (D\!-\!1) H^2} \Bigl[ -(4 y \!-\! y^2) \gamma'(y)
- (D\!-\!1) (2\!-\!y) \gamma(y)\Bigr] \; , \\
C(y) & \equiv & \frac1{4 (D\!-\!1) H^2} \Bigl[ (2 \!-\! y) \gamma'(y)
- (D \!-\! 1) \gamma(y)\Bigr] \; .
\end{eqnarray}
The function $\gamma(y)$ is,
\begin{eqnarray}
\lefteqn{\gamma(y) = - \Bigl(\frac{D\!-\!1}2\Bigr) \frac{H^2}{m^2}
\frac{H^{D-2}}{(4\pi)^{\frac{D}{2}}} \Bigg\{\!- \frac{\Gamma(D\!-\!1)}{
\Gamma(\frac{D}2 \!+\! 1)} \, {}_2 F_1\left(D\!-\!1,2;\frac{D}{2} \!+\! 1;
1 \!-\! \frac{y}4\right) } \nonumber \\
& & + \frac{\Gamma(\frac{D+1}2 \!+\! \nu) \Gamma(\frac{D+1}2 \!-\! \nu)}{
\Gamma(\frac{D}2 \!+\! 1)} \, {}_2 F_1\left(\frac{D\!+\!1}2 \!+\! \nu,
\frac{D\!+\!1}2 \!-\! \nu;\frac{D}2 \!+\! 1;1\!-\!\frac{y}4\right) \Bigg\} .
\qquad
\end{eqnarray}
Here the parameter $\nu$ is,
\begin{equation}
\nu \equiv \sqrt{ \Bigl(\frac{D\!-\!3}2\Bigr)^2 - \frac{m^2}{H^2} } \; .
\end{equation}

Because our work is limited to coincidence limits we require only the
integer powers in the Laurent expansion of $\gamma(y)$,
\begin{eqnarray}
\lefteqn{\gamma(y) = - \Bigl(\frac{D\!-\!1}2\Bigr) \frac{H^2}{m^2} 
\frac{H^{D-2}}{(4 \pi)^{\frac{D}{2}}} \sum_{n=0}^{\infty} \Biggl\{ -(n\!+\!1) 
\frac{\Gamma(n \!+\! D \!-\! 1)}{\Gamma(n \!+\! \frac{D}2 \!+\! 1)} 
\Bigl(\frac{y}4\Bigr)^n } \nonumber \\
& & \hspace{-.5cm} +\frac{\Gamma(\frac{D}2 \!-\! 1) \Gamma(2 \!-\! \frac{D}2)}{
\Gamma(n \!+\! \frac{D}2 \!+\! 1) n!} \frac{ \Gamma(n \!+\! \frac{D+1}2 \!+\! 
\nu) \Gamma(n \!+\! \frac{D+1}2 \!-\! \nu)}{ \Gamma(\frac12 \!+\! \nu)
\Gamma(\frac12 \!-\! \nu)} \Bigl(\frac{y}4\Bigr)^n + O\Bigl(y^{n-\frac{D}2+1}
\Bigr) \Biggr\} . \qquad
\end{eqnarray}
When $x^{\prime \mu} = x^{\mu}$, the function $y(x;x')$ vanishes. Because 
dimensional regularization ignores all $D$-dependent powers of zero, the
coincidence limits of $\gamma(y)$ and its derivatives derive from factors
of $y^0 = 1$. The two we require are,
\begin{eqnarray}
\gamma(0) = \Bigl(\frac{D\!-\!1}2\Bigr) \frac{H^2}{m^2} 
\frac{H^{D-2}}{(4 \pi)^{\frac{D}{2}}} \Biggl\{ \frac{\Gamma(D \!-\! 1)}{\Gamma(
\frac{D}2 \!+\! 1)}  - \frac{\Gamma(-\frac{D}2) \Gamma(\frac{D+1}2 \!+\! \nu) 
\Gamma(\frac{D+1}2 \!-\! \nu)}{ \Gamma(\frac12 \!+\! \nu) \Gamma(\frac12 
\!-\! \nu)} \Biggr\} , \; \label{gamcoin} \\
\gamma'(0) = \frac{(D\!-\!1)^2}{2 (D\!+\!2)} \frac{H^2}{m^2} 
\frac{H^{D-2}}{(4 \pi)^{\frac{D}{2}}} \Biggl\{ \frac{\Gamma(D \!-\!1)}{\Gamma(
\frac{D}2 \!+\! 1)} 
-\frac{\Gamma(- \frac{D}2) \Gamma(\frac{D+3}2 \!+\! \nu) \Gamma(\frac{D+3}2 
\!-\! \nu)}{ 2 (D\!-\!1) \Gamma(\frac12 \!+\! \nu) \Gamma(\frac12 \!-\! \nu)} 
\Biggr\} . \; \label{gampcoin}
\end{eqnarray}
And the two coincidence limits we need of the photon propagator are,
\begin{eqnarray}
\lefteqn{\lim_{x' \rightarrow x} \Bigl\langle x \Bigl\vert \frac{i}{{\cal 
D}^{\mu\nu}} \Bigr\vert x' \Bigr\rangle = \gamma(0) g_{\mu\nu} \; , 
\label{AAcoin} } \\
\lefteqn{\lim_{x' \rightarrow x} D_{\rho} D'_{\sigma} \Bigl\langle x \Bigl\vert 
\frac{i}{{\cal D}^{\mu\nu}} \Bigr\vert x' \Bigr\rangle = H^2 \Bigl[ -2 
\Bigl(\frac{D\!+\!1}{D\!-\!1}\Bigr) \gamma'(0) \!+\! \gamma(0)\Bigr] 
g_{\mu\nu} g_{\rho\sigma} } \nonumber \\
& & \hspace{2cm} + H^2 \Bigl[ \frac2{D\!-\!1} \gamma'(0) \Bigr] g_{\mu\rho}
g_{\nu\sigma} + H^2 \Bigl[ \frac2{D\!-\!1} \gamma'(0) \!-\! \gamma(0) \Bigr]
g_{\mu\sigma} g_{\nu\rho} \; . \qquad \label{Dcoin}
\end{eqnarray}
Here $D_{\rho}$ is the covariant derivative operator defined by $D_{\rho}
A_{\mu} \equiv A_{\mu ; \rho}$.

We can now integrate the vector potential out of the field strength and
scalar kinetic bilinears. First, recall that the ordinary derivatives in
the field strength tensor can be replaced by covariant derivatives,
\begin{equation}
F_{\mu\nu} \equiv \partial_{\mu} A_{\nu} - \partial_{\nu} A_{\mu} =
D_{\mu} A_{\nu} - D_{\nu} A_{\mu} \; .
\end{equation}
Combining this with relations (\ref{Dcoin}) and (\ref{gamcoin}-\ref{gampcoin}), 
and with our earlier insight about the $J^{\nu}$ terms dropping, we see that 
the leading logarithm form of the field strength bilinear is,
\begin{eqnarray}
\lefteqn{\Fint [dA_{\mu}] \delta\Bigl[\partial_{\mu}(\sqrt{-g} g^{\mu\nu} 
A_{\nu}) \Bigr] e^{i S[\varphi^*,\varphi,A]} \times F_{\alpha\beta}(x) 
F_{\rho\sigma}(x) } \nonumber \\
& & \hspace{-.3cm} \longrightarrow e^{i \Gamma[\varphi^*,\varphi]} \times
\lim_{x' \rightarrow x} \Biggl\{ D_{\alpha} D'_{\rho} \Bigl\langle x \Bigl\vert
\frac{i}{{\cal D}^{\beta \sigma}} \Bigr\vert x'\Bigr\rangle - D_{\alpha} 
D'_{\sigma} \Bigl\langle x \Bigl\vert \frac{i}{{\cal D}^{\beta \rho}} 
\Bigr\vert x'\Bigr\rangle \nonumber \\
& & \hspace{4.5cm} - D_{\beta} D'_{\rho} \Bigl\langle x \Bigl\vert \frac{i}{
{\cal D}^{\alpha \sigma}} \Bigr\vert x'\Bigr\rangle + D_{\beta} D'_{\sigma} 
\Bigl\langle x \Bigl\vert \frac{i}{{\cal D}^{\alpha \rho}} \Bigr\vert 
x'\Bigr\rangle \Biggr\} , \qquad \\
& & \hspace{-.3cm} = e^{i \Gamma[\varphi^*,\varphi]} \times 
\Bigl(g_{\alpha\rho} g_{\beta \sigma} - g_{\alpha\sigma} g_{\beta\rho} \Bigr) 
H^2 \Bigl[-4 \Bigl(\frac{D \!+\!2}{D\!-\!1}\Bigr) \gamma'(0) + 
4 \gamma(0)\Bigr] \; , \\
& & \hspace{-.3cm} = e^{i \Gamma[\varphi^*,\varphi]} \times 
\Bigl(g_{\alpha\rho} g_{\beta \sigma} - g_{\alpha\sigma} g_{\beta\rho} \Bigr) 
\frac{H^D}{(4\pi)^{\frac{D}2}} \frac{\Gamma(-\frac{D}2) \Gamma(\frac{D+1}2 
\!+\! \nu) \Gamma(\frac{D+1}2 \!-\! \nu)}{\Gamma(\frac12 \!+\! \nu)
\Gamma(\frac12 \!-\! \nu)} \; . \qquad \label{Fexp}
\end{eqnarray}
The analogous result for the scalar kinetic bilinear is,
\begin{eqnarray}
\lefteqn{e^{-i \Gamma[\varphi^*,\varphi]} \times \Fint [dA_{\mu}] 
\delta\Bigl[\partial_{\mu}(\sqrt{-g} g^{\mu\nu} A_{\nu}) \Bigr] 
e^{i S[\varphi^*,\varphi,A]} \times \Bigr(D_{\rho} \varphi(x) \Bigr)^* 
D_{\sigma} \varphi(x) } \nonumber \\
& & \longrightarrow e^2 
\varphi^*(x) \varphi(x) \times \Bigl\langle x \Bigl\vert \frac{i}{{\cal D}^{
\rho\sigma}} \Bigr\vert x \Bigr\rangle \; , \\
& & = e^2 \varphi^* 
\varphi \times g_{\rho\sigma} \gamma(0) \; , \\
& & = g_{\rho\sigma} 
\Bigl(\frac{D\!-\!1}4\Bigr) \frac{H^D}{(4\pi)^{\frac{D}2}} \Biggl\{
\frac{\Gamma(D\!-\!1)}{\Gamma(\frac{D}2 \!+\!1)} - \frac{\Gamma(-\frac{D}2) 
\Gamma(\frac{D+1}2 \!+\! \nu) \Gamma(\frac{D+1}2 \!-\! \nu)}{\Gamma(\frac12 
\!+\! \nu) \Gamma(\frac12 \!-\! \nu)} \Biggr\} . \qquad \label{phiexp}
\end{eqnarray}

Before computing the stress tensor we should comment on the explicit 
perturbative computations which have been done to check the field strength
bilinear (\ref{Fexp}) and the scalar kinetic bilinear (\ref{phiexp}). Of
course there is no way to check the nonperturbative information these
expressions contain! However, we can compare against explicit one and two
loop computations by expanding the parameter $\nu$ in powers of $m^2 =
2 e^2 \varphi^* \varphi$,
\begin{equation}
\nu = \sqrt{\Bigl(\frac{D\!-\!3}2\Bigr)^2 - \frac{m^2}{H^2}} \equiv
\Bigl(\frac{D\!-\!3}2\Bigr) - \Delta \nu = \Bigl(\frac{D\!-\!3}2\Bigr)
- \frac1{D\!-\!3} \frac{m^2}{H^2} + O\Bigl(\frac{m^4}{H^4}\Bigr) \; .
\end{equation}
From (\ref{Fexp}) we see that the leading logarithm result for the
field strength bilinear should be,
\begin{eqnarray}
\lefteqn{ \Bigl\langle F_{\alpha\beta}(x) F_{\rho\sigma}(x) 
\Bigr\rangle_{\rm lead\ log} } \nonumber \\
& & = \Bigl(g_{\alpha\rho} g_{\beta\sigma} - g_{\alpha\sigma} g_{\beta\rho}
\Bigr) \frac{H^D}{(4\pi)^{\frac{D}2}} \Gamma\Bigl(-\frac{D}2\Bigr) 
\Biggl\langle \frac{\Gamma(D \!-\! 1 \!-\! \Delta \nu) \Gamma(2 \!+\! \Delta 
\nu)}{\Gamma(\frac{D}2 \!-\! 1 \!-\! \Delta \nu) \Gamma(2 \!-\! \frac{D}2 
\!+\! \Delta \nu)} \Biggr\rangle \; , \qquad \\
& & = \Bigl(g_{\alpha \rho} g_{\beta \sigma} - g_{\alpha \sigma} g_{\beta\rho}
\Bigr) \frac{H^D}{(4\pi)^{\frac{D}2}} \frac{\Gamma(D\!-\!1)}{\Gamma(\frac{D}2
\!+\! 1)} \Biggl\{1 + \Biggl[-\psi(D\!-\!1) \nonumber \\
& & \hspace{2.5cm} + \psi(2) + \psi\Bigl(\frac{D}2 \!-\! 1 \Bigr) - \psi\Bigl(2 
\!-\! \frac{D}2\Bigr) \Biggr] \Bigl\langle \Delta \nu \Bigr\rangle + 
\Bigl\langle O(\Delta \nu^2) \Bigr\rangle \Biggr\} . \qquad
\end{eqnarray}
Now substitute the leading stochastic result using relation 
(\ref{arbD}),
\begin{eqnarray}
\Bigl\langle \Delta \nu\Bigr\rangle & = & \frac{2 e^2}{(D\!-\!3) H^2}
\Bigl\langle \varphi^* \varphi \Bigr\rangle + O\Bigl(e^4 \Bigl\langle 
(\varphi^* \varphi)^2 \Bigr\rangle \Bigr) \; , \\
& \longrightarrow & \frac{4 e^2 H^{D-4}}{(4 \pi)^{\frac{D}2}} \frac{\Gamma(D
\!-\! 1)}{(D \!-\!3) \Gamma(\frac{D}2)} \ln(a) + O\Bigl(e^4 \ln^2(a)\Bigr) \; .
\end{eqnarray}
Because the two loop result is divergent we report only the divergent part,
\begin{eqnarray}
\lefteqn{\Bigl\langle F_{\alpha\beta}(x) F_{\rho\sigma}(x) \Bigr\rangle_{\rm 
lead\ log} } \nonumber \\
& & = \Bigl(g_{\alpha \rho} g_{\beta \sigma} - g_{\alpha \sigma} g_{\beta\rho}
\Bigr) \frac{H^4}{16 \pi^2} \Biggl\{1 - \frac1{D\!-\!4} \times \frac{e^2}{
\pi^2} \ln(a) + O\Bigl(e^4 \ln^2(a)\Bigr) \Biggr\} . \qquad
\end{eqnarray}
The one loop (order one) result is trivial and is not, in any case, a check
of the technique because it contains no infrared logarithm. The two loop 
(order $e^2$) result agrees with the diagram given in Fig.~\ref{FF} but we 
have not yet proved that the other two loop diagrams fail to contribute 
divergent leading logarithms.

The scalar kinetic bilinear has been more thoroughly checked. Our 
sto\-chas\-tic prediction for it is,
\begin{eqnarray}
\lefteqn{ \Bigl(D_{\rho} \varphi(x)\Bigr)^* D_{\sigma} \varphi(x) 
\Bigr\rangle_{\rm lead\ log} = g_{\rho\sigma} \Bigl(\frac{D\!-\!1}4\Bigr)
\frac{H^D}{(4\pi)^{\frac{D}2}} } \nonumber \\
& & \hspace{2cm} \times \Biggl\{ \frac{\Gamma(D\!-\!1)}{\Gamma(\frac{D}2 
\!+\! 1)} - \Gamma\Bigl(-\frac{D}2\Bigr) \Biggl\langle \frac{\Gamma(D \!-\! 
1 \!-\! \Delta \nu) \Gamma(2 \!+\! \Delta \nu)}{\Gamma(\frac{D}2 \!-\! 1 
\!-\! \Delta \nu) \Gamma(2 \!-\! \frac{D}2 \!+\! \Delta \nu)} \Biggr\rangle 
\!\Biggr\} , \qquad \\
& & = g_{\rho\sigma} \Bigl(\frac{D\!-\!1}4\Bigr) \frac{H^D}{(4\pi)^{\frac{D}2}}
\frac{\Gamma(D\!-\!1)}{\Gamma(\frac{D}2 \!+\! 1)} \Biggl\{
\Biggl[\psi(D\!-\!1) \nonumber \\
& & \hspace{2cm} - \psi(2) - \psi\Bigl(\frac{D}2 \!-\! 1 \Bigr) + \psi\Bigl(2 
\!-\! \frac{D}2\Bigr) \Biggr] \Bigl\langle \Delta \nu \Bigr\rangle + 
\Bigl\langle O(\Delta \nu^2) \Bigr\rangle \Biggr\} , \qquad \\
& & \longrightarrow g_{\rho \sigma} \frac{e^2 H^{2D-4}}{(4\pi)^{D}} 
\frac{\Gamma(D\!-\!1) \Gamma(D)}{(D\!-\!3) \Gamma(\frac{D}2) \Gamma(\frac{D}2
\!+\!1)} \Biggl[ \psi\Bigl(2 \!-\! \frac{D}2\Bigr) \nonumber \\
& & \hspace{2cm} - \psi\Bigl(\frac{D}2 \!-\! 1) + \psi(D \!-\! 1) - \psi(2)
\Biggr] \ln(a) + O\Bigl(e^4 \ln^2(a)\Bigr) . \qquad
\end{eqnarray}
This agrees exactly, and for arbitrary dimension $D$, with the infrared 
logarithm in equation (146) of our recent, explicit two loop computation 
of the scalar kinetic bilinear \cite{PTsW}.

We can now assemble the various constituents of the stress tensor. From
(\ref{ctms}) and (\ref{delxi}) we that the conformal counterterm contributes,
\begin{equation}
-(D \!-\! 1) (D \!-\! 2) \delta \xi H^2 \varphi^* \varphi g_{\mu\nu} = 
- g_{\mu\nu} \times (D \!-\! 1) (D \!-\! 2) \frac{H^D}{(4\pi)^{\frac{D}2}}
\Biggl\{ \frac{z}{\epsilon} + \frac{\gamma z}2 + O(\epsilon) \Biggr\} .
\label{xicon}
\end{equation}
Recall that $z \equiv e^2 \varphi^* \varphi/H^2$ and $\epsilon \equiv 
4 - D$. We shall keep the same form as (\ref{xicon}) for the each of the
four terms. From (\ref{ctms}) and (\ref{dellam}) the $(\varphi^* \varphi)^2$
counterterm gives,
\begin{equation}
-\frac{\delta \lambda}{4} (\varphi^* \varphi)^2 g_{\mu\nu} = - g_{\mu\nu} 
\times (D \!-\! 1) (D \!-\! 2) \frac{H^D}{(4\pi)^{\frac{D}2}}
\Biggl\{ \frac{z^2}{\epsilon} + \frac{\gamma \!-\! 1}2 z^2 + O(\epsilon) 
\Biggr\} . \label{lamcon}
\end{equation}
The field strength contribution results from combining (\ref{fullT}) with
(\ref{Fexp}), and it turns out to be finite,
\begin{eqnarray}
\lefteqn{\Bigl[\delta^{\alpha}_{\mu} \delta^{\rho}_{\nu} \!-\! \frac14 
g_{\mu\nu} g^{\alpha\rho} \Bigr] g^{\beta\sigma} \times
\Bigl(g_{\alpha\rho} g_{\beta \sigma}- g_{\alpha\sigma} 
g_{\beta\rho} \Bigr) } \nonumber \\
& & \hspace{4cm} \times \frac{H^D}{(4\pi)^{\frac{D}2}} \frac{\Gamma(-\frac{D}2) 
\Gamma(\frac{D+1}2 \!+\! \nu) \Gamma(\frac{D+1}2 \!-\! \nu)}{\Gamma(\frac12 
\!+\! \nu) \Gamma(\frac12 \!-\! \nu)} \; , \qquad \\
& & = \frac14 (D \!-\! 4) (D \!-\! 1) g_{\mu\nu} 
\frac{H^D}{(4\pi)^{\frac{D}2}} \frac{\Gamma(-\frac{D}2) 
\Gamma(\frac{D+1}2 \!+\! \nu) \Gamma(\frac{D+1}2 \!-\! \nu)}{\Gamma(\frac12 
\!+\! \nu) \Gamma(\frac12 \!-\! \nu)} \; , \\
& & = - g_{\mu\nu} \times (D\!-\!1) (D \!-\! 2) \frac{H^D}{(4\pi)^{\frac{D}2}}
\Biggl\{ \frac0{\epsilon} - \frac12 (z \!+\! z^2) + O(\epsilon) \Biggr\} . 
\label{Fcon}
\end{eqnarray}
The contribution of the scalar kinetic term comes from substituting 
(\ref{phiexp}) in (\ref{fullT}) and then making use of the same expansion
(\ref{gamexp}) as for the effective potential,
\begin{eqnarray}
\lefteqn{\Bigl[\delta_{\mu}^{\rho} \delta_{\nu}^{\sigma} \!+\!
\delta_{\mu}^{\sigma} \delta_{\nu}^{\rho} \!-\! g_{\mu\nu} g^{\rho\sigma}
\Bigr] \times g_{\rho\sigma} } \nonumber \\
& & \hspace{1cm} \times 
\Bigl(\frac{D\!-\!1}4\Bigr) \frac{H^D}{(4\pi)^{\frac{D}2}} \Biggl\{
\frac{\Gamma(D\!-\!1)}{\Gamma(\frac{D}2 \!+\!1)} - \frac{\Gamma(-\frac{D}2) 
\Gamma(\frac{D+1}2 \!+\! \nu) \Gamma(\frac{D+1}2 \!-\! \nu)}{\Gamma(\frac12 
\!+\! \nu) \Gamma(\frac12 \!-\! \nu)} \Biggr\} , \qquad \\
& & \hspace{-.6cm} = -\frac{(D\!-\!2) (D\!-\!1)}4 \frac{g_{\mu\nu} H^D}{
(4\pi)^{\frac{D}2}} \Biggl\{\!\frac{\Gamma(D\!-\!1)}{\Gamma(\frac{D}2 \!+\!1)} 
- \frac{\Gamma(-\frac{D}2) \Gamma(\frac{D+1}2 \!+\! \nu) \Gamma(\frac{D+1}2 
\!-\! \nu)}{\Gamma(\frac12 \!+\! \nu) \Gamma(\frac12 \!-\! \nu)} \!\Biggr\} 
, \qquad \\
& & \hspace{-.6cm} = - g_{\mu\nu} \times (D\!-\!1) (D \!-\! 2) \frac{H^D}{
(4\pi)^{\frac{D}2}} \Biggl\{ -\Bigl(\frac{z \!+\! z^2}{\epsilon}\Bigr) + 
\Bigl(-\frac12 + \frac{\gamma}2\Bigr) z + \Bigl(-\frac34 + \frac{\gamma}2\Bigr)
z^2 \nonumber \\
& & \hspace{4cm} + \frac12 (z \!+\! z^2) \Bigl[\psi\Bigl(\frac32 \!+\! 
\nu\Bigr) + \psi\Bigl(\frac32 \!-\! \nu\Bigr)\Bigr] + O(\epsilon) \Biggr\} . 
\label{phicon}
\end{eqnarray}
Of course the divergences in (\ref{xicon}) and (\ref{lamcon}) cancel those
in (\ref{phicon}), at which point we can take $D=4$. The final result has
the form $-g_{\mu\nu} V_s(z)$ where,
\begin{eqnarray}
\lefteqn{V_{\rm s} = \frac{3 H^4}{8 \pi^2} \Biggl\{ (-1 + \gamma) z
+ \Bigl(-\frac74 + \gamma\Bigr) z^2 } \nonumber \\
& & \hspace{2cm} + \frac12 z (1 + z) \Biggl[ \psi\Bigl(\frac32 + 
\frac12 \sqrt{1 - 8 z}\Bigr) + \psi\Bigl(\frac32 -\frac12 
\sqrt{1-8z}\Bigr)\Biggr] \Biggr\} . \qquad \label{Vs}
\end{eqnarray}

\begin{figure}
\centerline{\epsfig{file=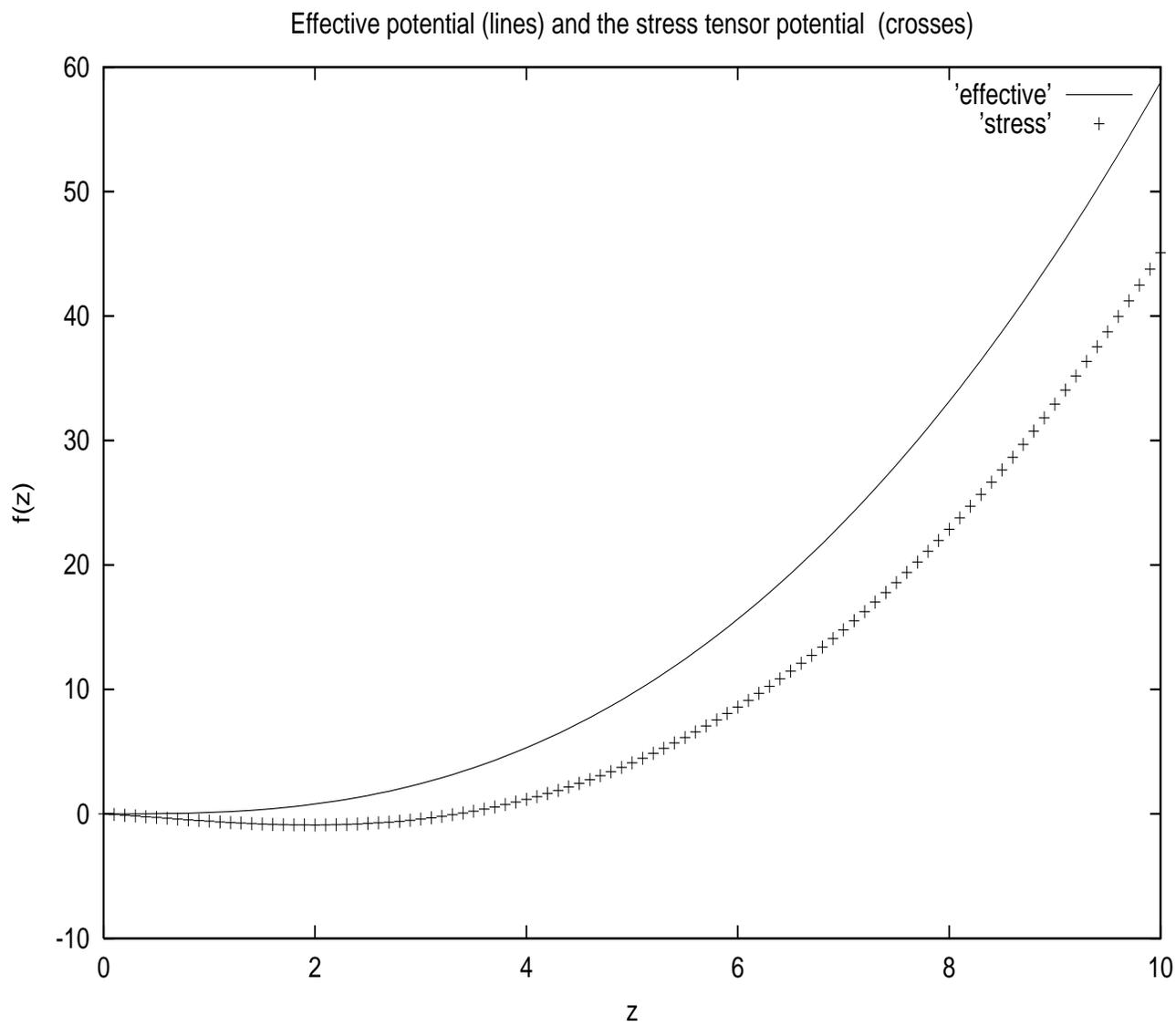,height=6.9in,width=6in,angle=180}}
\caption{Effective potential $V_{\rm eff} \equiv \frac{3 H^2}{8 \pi^2}
f(z)$ (in solid lines) versus stress potential $V_{\rm s} \equiv
\frac{3 H^2}{8 \pi^2} f_{\rm s}(z)$ (crosses).}
\label{fullstress}
\end{figure}

It is apparent from Fig.~\ref{fullstress} that the stress tensor potential 
$V_{\rm s}(\varphi^* \varphi)$ does not quite agree with the effective 
potential $V_{\rm eff}(\varphi^* \varphi)$. This same sort of disagreement
was also noted in the recent leading logarithm solution of Yukawa theory
\cite{MW1}. In both models the difference arises because the two potentials 
describe different physical processes: $V_{\rm eff}$ controls the scalar's
evolution whereas $V_{\rm s}$ controls the gravitational back-reaction.
The two are distinct because almost all the factors of $H^2$ in $V_{\rm eff}$
are $R/12$ for a general metric, and this changes the stress tensor, even in
de Sitter background and at leading logarithm order. 

To see the point, consider a contribution to the matter Lagrangian 
of the form,
\begin{equation}
\Delta {\cal L} = -F(R) \sqrt{-g} \; .
\end{equation}
The corresponding contribution to the stress tensor is,
\begin{eqnarray}
\Delta T_{\mu\nu} & \equiv & -\frac2{\sqrt{-g}} \, \frac{\delta \Delta S}{
\delta g^{\mu\nu}} \; , \\
& = & 2 R_{\mu\nu} F'(R) - g_{\mu\nu} F(R) + 2 g_{\mu\nu} F(R)^{;\rho}_{~~\rho}
- 2 F(R)_{;\mu\nu} \; .
\end{eqnarray}
In $D=4$ de Sitter background $R_{\mu\nu} = g_{\mu\nu} R/4$, and we can 
ignore the derivative terms at leading logarithm order,
\begin{equation}
\Delta T_{\mu\nu} \longrightarrow -g_{\mu\nu} \Bigl[F(R) - \frac12 R F'(R)
\Bigr] \; .
\end{equation}
If all the factors of $H^2$ in $V_{\rm eff}$ were $R/12$ for a general metric,
then for de Sitter background we would have,
\begin{equation}
F(R) = \frac3{8 \pi^2} \Bigl(\frac{R}{12}\Bigr)^2 f\Bigl(\frac{12 e^2
\varphi^* \varphi}{R}\Bigr) \; \Longrightarrow \; \Bigl[F(R) - \frac12 R 
F'(R) \Bigr]_{\rm dS} = \frac{3 H^4}{8 \pi^2} \times \frac12 z f'(z) 
\; . \label{putative}
\end{equation}

The actual relation between $V_{\rm eff}$ and $V_{\rm s}$ is tantalizingly
close to (\ref{putative}). If we extract a factor of $3 H^4/8\pi^2$ from 
each potential,
\begin{equation}
V_{\rm eff}(\varphi^* \varphi) \equiv \frac{3 H^4}{8 \pi^2} \, f(z) \qquad
{\rm and} \qquad V_{\rm s}(\varphi^* \varphi) \equiv \frac{3 H^4}{8 \pi^2} 
\, f_{\rm s}(z) \; , \label{fdef}
\end{equation}
then comparison of expressions (\ref{Veff}) and (\ref{Vs}) reveals the
following relation between the two dimensionless functions of $z$,
\begin{equation}
f_{\rm s}(z) = \frac12 z f'(z) - \frac12 z - \frac14 z^2 \; . \label{frel}
\end{equation}
The reason for the extra contribution of $-z/2 - z^2/4$ is that a small
portion of the $H^2$ dependence in $V_{\rm eff}$ is really the constant
$\Lambda/3$, rather than $R/12$. These are the finite factors of $\ln(H^2)$ 
which derive from our counterterms (\ref{delxi}-\ref{dellam}) through the
expansion,
\begin{equation}
\frac{H^{D-4}}{4 \!-\! D} = \frac1{4 \!-\! D} - \frac12 \ln(H^2) + 
O(4 \!-\! D) \; .
\end{equation}
If we regard the arbitrary generalization of these factors of $\ln(H^2)$ 
as $\ln(\Lambda/3)$, rather than $\ln(R/12)$, it corresponds to adding
the following term to $F(R)$,
\begin{equation}
\Delta F(R) = \frac3{8\pi^2} \Bigl(\frac{R}{12}\Bigr)^2 \Biggl[
\frac{12 e^2 \varphi^* \varphi}{R} + \frac12 \Bigl( \frac{12 e^2 \varphi^* 
\varphi}{R}\Bigr)^2 \Biggr] \ln\Bigl(\frac{R}{4 \Lambda}\Bigr) \; .
\end{equation}
Because the logarithm vanishes for $R = 12 H^2$, the only change in
$F - R F'/2$ in de Sitter background is precisely the required deficit term,
\begin{equation}
\Bigl[\Delta F(R) - \frac12 R \Delta F'(R) \Bigr]_{\rm dS} = \frac{3 H^4}{8 
\pi^2} \Bigl\{-\frac12 z - \frac14 z^2\Bigr\} \; .
\end{equation}

\section{Nonperturbative Predictions}

It would be silly to stop without exploiting the formalism we have developed
to answer nonperturbative questions about SQED. One would like to know:
\begin{enumerate}
\item{How large does the scalar field strength become?}
\item{What is the asymptotic late time value of the photon mass?}
\item{Does the scalar remain light?}
\item{Does the vacuum energy increase or decrease, and by how much?}
\item{What becomes of the electric and magnetic field strengths?}
\end{enumerate}
Answering these questions is the task of this section. We begin by making
the trivial generalization of Starobinski\u{\i}'s formalism from a real 
scalar to a complex one. We then exploit the results of the previous two
sections to compute explicit answers to each of the five questions.

The stochastic formalism of subsections 2.4 and 2.5 has a straightforward
generalization to a complex scalar. One simply decomposes the complex
field into two real scalars in the usual way,
\begin{equation}
\varphi(x) \equiv \frac1{\sqrt{2}} \Bigl( \varphi_1(x) + i \varphi_2(x) 
\Bigr) \; .
\end{equation}
Now suppose expectation values of the quantum fields $\varphi_i(x)$ agree,
at leading logarithm order, with those of the stochastic random variables
$\Phi_i(x)$ which obey the Langevin equations,
\begin{equation}
\dot{\Phi}_i = \dot{\Phi}_{i0} - \frac1{3 H} \frac{\partial V_{\rm eff}}{
\partial \Phi_i} \; .
\end{equation}
The fields $\Phi_{i0}(t,\vec{x})$ are independent sources of Gaussian 
white noise,
\begin{equation}
\Bigl\langle \Phi_{i0}(t,\vec{x}) \Phi_{j0}(t',\vec{x}) \Bigr\rangle
= \frac{H^3}{4 \pi^2} \delta(t-t') \delta_{ij} \; .
\end{equation}
Then the expectation value of any function of the $\Phi_i$ is given in
terms of a probability density $\varrho(t,\phi_1,\phi_2)$,
\begin{equation}
\Bigl\langle F\Bigl[\Phi_1(t,\vec{x}),\Phi_2(t,\vec{x}) \Bigr] \Bigr\rangle
= \int_{-\infty}^{\infty} \!\!\! d\phi_1 \int_{-\infty}^{\infty} \!\!\! 
d\phi_2 \, \varrho(t,\phi_1,\phi_2) F(\phi_1,\phi_2) \; .
\end{equation}
This probability density obeys the Fokker-Planck equation,
\begin{equation}
\dot{\varrho} = \frac1{3H} \sum_{i=1}^2 \frac{\partial}{\partial \phi_i}
\Bigl\{ \frac{\partial V_{\rm eff}}{\partial \phi_i} \varrho \Bigr\} + \frac12
\sum_{i=1}^2 \frac{\partial^2}{\partial \phi_i^2} \Bigl\{ \frac{H^3}{4\pi^2}
\varrho \Bigr\} \; .
\end{equation}

We now exploit the fact that the effective potential depends upon the
$\varphi_i$ only through the combination,
\begin{equation}
\varphi^* \varphi = \frac12 \Bigl( \varphi_1^2 + \varphi_2^2\Bigr) \; .
\end{equation}
It follows that the probability density has the same form and that we
can write the Fokker-Planck equation as,
\begin{equation}
\dot{\varrho} = \frac1{3 H} \sum_{i=1}^2 \frac{\partial}{\partial \phi_i}
\Bigl\{ \phi_i V'_{\rm eff} \varrho \Bigr\} + \frac{H^3}{8 \pi^2}
\sum_{i=1}^2 \frac{\partial}{\partial \phi_i} \Bigl\{ \phi_i \varrho' \Bigr\}
\; . \label{ourFP}
\end{equation}
Here a prime denotes differentiation with respect to the variable $\phi^*
\phi \equiv (\phi_1^2 + \phi_2^2)/2$.

It is apparent from the large field expansion (\ref{largephi}), and from
Fig.~\ref{full}, that $V_{\rm eff}$ is bounded below. We can therefore make 
the ansatz of Starobinski\u{\i} and Yokoyama \cite{SY} that the probability
density approaches a time independent form at late times,
\begin{equation}
\lim_{t \rightarrow \infty} \varrho(t,\phi_1,\phi_2) = \varrho_{\infty}(\phi^* 
\phi) \; .
\end{equation}
Substituting this in our Fokeer-Planck equation (\ref{ourFP}) and making a 
few simple inferences implies,
\begin{equation}
\varrho_{\infty}(\phi^*\phi) V'_{\rm eff}(\phi^*\phi) = -\frac{3 H^4}{8\pi^2}
\varrho'_{\infty}(\phi^* \phi) \quad \Longrightarrow \quad 
\varrho_{\infty}(\phi^* \phi) = N e^{-\frac{8\pi^2}{3 H^4} 
V_{\rm eff}(\phi^*\phi)} \; .
\end{equation}
The asymptotic probability density can be more simply expressed in terms
of the function $f(z)$ introduced in equation (\ref{fdef}),
\begin{equation}
\varrho_{\infty}(\phi^*\phi) = N e^{-f(z)} \; ,
\end{equation}
where $z \equiv e^2 \phi^* \phi/H^2$ and,
\begin{eqnarray}
\lefteqn{f(z) = (-1 + 2 \gamma) z + (-\frac32 + \gamma) z^2 } \nonumber \\
& & \hspace{2cm} + \int_0^z \!\!\! dx \, (1 \!+\! x) \Biggl[ \psi\Bigl(
\frac32 \!+\! \frac12 \sqrt{1 \!-\! 8 x}\Bigr) + \psi\Bigl(\frac32 \!-\!
\frac12 \sqrt{1 \!-\! 8x}\Bigr)\Biggr] . \qquad \label{f(z)}
\end{eqnarray}
Hence the late time limit of any function of the operator $\varphi^*(x) 
\varphi(x)$ can be reduced to an ordinary integral,
\begin{eqnarray}
\lim_{t \rightarrow \infty} \Bigl\langle F\Bigl(\varphi^*(x) \varphi(x)\Bigr)
\Bigr\rangle & = & \int_{-\infty}^{\infty} \!\!\! d\phi_1 
\int_{-\infty}^{\infty} \!\!\! d\phi_2 \, F(\phi^* \phi\Bigr) \, 
\varrho_{\infty}(\phi^*\phi) \; , \qquad \\
& = & 2 \pi N \int_0^{\infty} \!\!\! dz \, z F\Bigl(\frac{H^2 z}{e^2}\Bigr) \,
e^{-f(z)} \; . \qquad \label{arbexp}
\end{eqnarray}

\begin{table}

\vbox{\tabskip=0pt \offinterlineskip
\def\tablerule{\noalign{\hrule}}
\halign to370pt {\strut#& \vrule#\tabskip=1em plus2em& 
\hfil#\hfil& \vrule#& \hfil#\hfil& \vrule#\tabskip=0pt\cr
\tablerule
\omit&height4pt&\omit&&\omit&\cr
&&$\!\!\!\!{\rm Operator}\!\!\!\!$ && $\!\!\!\! {\rm Expectation\ Value}
\!\!\!\!$ & \cr
\omit&height4pt&\omit&&\omit&\cr
\tablerule
\omit&height2pt&\omit&&\omit&\cr
&& $\varphi^* \varphi$ && $1.6495 \times H^2/e^2$ & \cr
\omit&height2pt&\omit&&\omit&\cr
\tablerule
\omit&height2pt&\omit&&\omit&\cr
&& $(\varphi^* \varphi)^2$ && $3.3213 \times H^4/e^4$ & \cr
\omit&height2pt&\omit&&\omit&\cr
\tablerule
\omit&height2pt&\omit&&\omit&\cr
&& $(\varphi^* \varphi)^3$ && $7.6308 \times H^6/e^6$ & \cr
\omit&height2pt&\omit&&\omit&\cr
\tablerule
\omit&height2pt&\omit&&\omit&\cr
&& $M^2_{\gamma} \equiv 2 e^2 \varphi^* \varphi$ && $3.2991 \times 
H^2$ & \cr
\omit&height2pt&\omit&&\omit&\cr
\tablerule
\omit&height2pt&\omit&&\omit&\cr
&& $M^2_{\varphi} \equiv V'_{\rm eff}(\varphi^* \varphi)$ && $.8961 \times 
3 e^2 H^2/8 \pi^2$ & \cr
\omit&height2pt&\omit&&\omit&\cr
\tablerule
\omit&height2pt&\omit&&\omit&\cr
&& $V_{\rm eff}(\varphi^* \varphi)$ && $.7223 \times 3 H^4/8 \pi^2$ & \cr
\omit&height2pt&\omit&&\omit&\cr
\tablerule
\omit&height2pt&\omit&&\omit&\cr
&& $V_{\rm s}(\varphi^* \varphi)$ && $-.6551 \times 3 H^4/8 \pi^2$ & \cr
\omit&height2pt&\omit&&\omit&\cr
\tablerule
\omit&height2pt&\omit&&\omit&\cr
&& $(F_{\mu\nu} F_{\rho\sigma})_{\rm fin}$ && $-9.5246 \times H^4/8 \pi^2 
\; (g_{\mu\rho} g_{\nu\sigma} - g_{\mu\sigma} g_{\nu\rho})$ & \cr
\omit&height2pt&\omit&&\omit&\cr
\tablerule}}

\caption{Late time limits of expectation values of some important operators.}

\label{lims}

\end{table}

It is of course impossible to obtain analytic expressions for integrals of
the form (\ref{arbexp}) with the function $f(z)$ in (\ref{f(z)}). However,
it is nothing these days to evaluate such integrals numerically. We have done
this using the ``NIntegrate'' function of Mathematica \cite{Wolfram}. The
normalization factor is about,
\begin{equation}
2 \pi N \equiv \Biggl[ \int_0^{\infty} \!\!\! dz \, z e^{-f(z)} \Biggr]^{-1} 
\approx \frac1{2.16603} \; .
\end{equation}
The scalar reaches a nonperturbatively large field strength,
\begin{equation}
\lim_{t \rightarrow \infty} \Bigl\langle \varphi^*(x) \varphi(x) \Bigr\rangle
= \frac{H^2}{e^2} \times 2\pi N \int_0^{\infty} \!\!\! dz \, z^2 e^{-f(z)}
\approx \frac{H^2}{e^2} \times \frac{3.57293}{2.16603} \; . \label{fstrength}
\end{equation}
The photon mass-squared is $M^2_{\gamma} \equiv 2 e^2 \varphi^* \varphi$, so
(\ref{fstrength}) means that it reaches the asymptotic value,
\begin{equation}
\lim_{t \rightarrow \infty} M^2_{\gamma} \approx 3.2991 \times H^2 \; .
\end{equation}
This is explicit, nonperturbative confirmation of the conjecture by Davis,
Dimopoulos, Prokopec and T\"ornkvist \cite{DDPT,DPTD} that inflation induces
a nonzero mass photon mass. Indeed, the nonperturbative result is about a
hundred times larger than one loop computations \cite{PTW2,PTW,PW1,PW3,PP1,PP2}.

Because inflationary particle production would be quenched if the scalar
were to develop a mass comparable to the Hubble parameter, it is important
to check that the scalar remains light. The scalar mass-squared is the 
derivative of the effective potential,
\begin{equation}
M^2_{\varphi} = V'_{\rm eff}(\varphi^* \varphi) \; .
\end{equation}
Our result for its asymptotic value is,
\begin{equation}
\lim_{t \rightarrow \infty} \Bigl\langle M^2_{\varphi} \Bigr\rangle =
\frac{3 e^2 H^2}{8 \pi^2} \times 2 \pi N \int_0^{\infty} \!\!\! dz \, z
f'(z) e^{-f(z)} \approx \frac{3 e^2 H^2}{8 \pi^2} \times 
\frac{1.94103}{2.16603} \; .
\end{equation}
Therefore the scalar is always light compared to the Hubble scale, and
the approximation of treating it stochastically with the massless mode
functions is justified.

\begin{figure}
\centerline{\epsfig{file=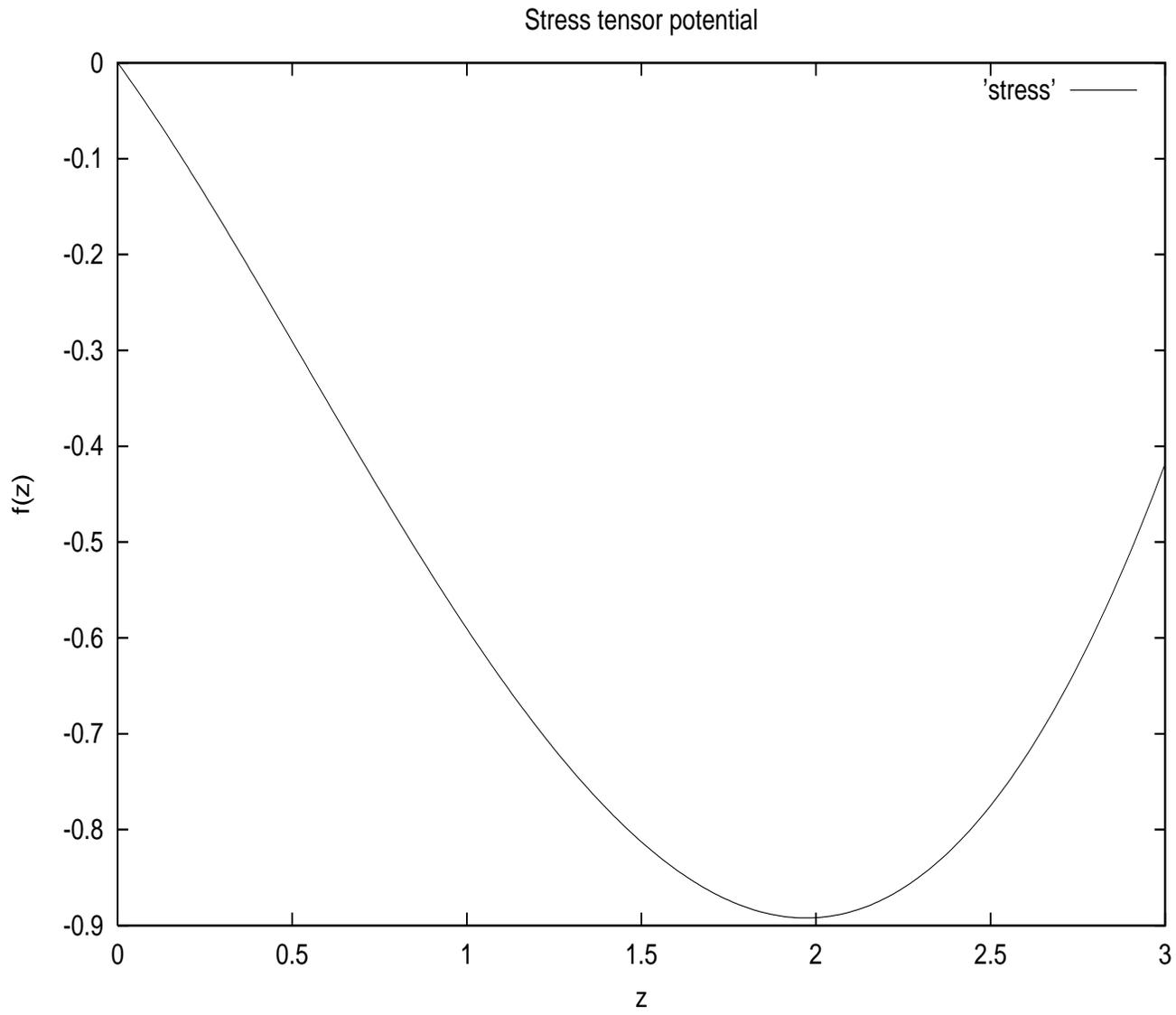,height=7in,width=6in,angle=180}}
\caption{Stress tensor potential $V_s \equiv \frac{3 H^4}{8\pi^2} 
f_{\rm s}(z)$ showing the minimum.}
\label{medstress}
\end{figure}

Table~\ref{lims} summarizes our results for late time limits of various
operators. One operator of particular interest is the stress tensor potential,
$V_{\rm s}(\varphi^*\varphi)$, given in equation (\ref{Vs}). In view of 
relations (\ref{fdef}-\ref{frel}) we can express its late time limit as,
\begin{eqnarray}
\lim_{t \rightarrow \infty} \Bigl\langle V_{\rm s}(\varphi^* \varphi) 
\Bigr\rangle & = & \frac{3 H^4}{8 \pi^2} \times 2 \pi N \int_0^{\infty} 
\!\!\! dz \, z \Bigl(\frac{z}2 f'(z) - \frac{z}2 - \frac{z^2}4\Bigr) 
e^{-f(z)} \; , \qquad \\
& \approx & \frac{3 H^4}{8 \pi^2} \times -\frac{1.41898}{2.16603} \; .
\end{eqnarray}
That the surprising sign is correct can be seen from Fig.~\ref{medstress}
which gives an expanded view of the stress tensor potential. It should be
noted that the sign is due to the two negative terms in the $z$ integrand.
With just the first term the result would be,
\begin{equation}
\frac{3 H^4}{8 \pi^2} \times 2 \pi N \int_0^{\infty} \!\!\! dz \, 
\frac{z^2}2 f'(z) e^{-f(z)} = + \frac{3 H^4}{8 \pi^2} \; .
\end{equation}
The physical interpretation may be that inflationary particle production 
polarizes the vacuum, which lowers the energy of a charged particle in the 
medium provided the charge density is not too large. 

Note from Table~\ref{lims} that the expectation value of the effective 
potential is positive. As explained at the end of the previous section, 
this is no contradiction with our result for $V_{\rm s}$ because the two 
potentials answer slightly different physical questions. $V_{\rm eff}$ 
controls the evolution of the scalar field strength whereas $V_{\rm s}$ 
controls the gravitational response. In particular, our result for $V_{\rm s}$
implies that inflationary particle production induces a small fractional 
reduction of the asymptotic expansion rate,
\begin{eqnarray}
3 H^2_{\infty} & \approx & 3 H^2 - 8 \pi G \times \frac{3 H^4}{8 \pi^2} 
\times .6551 \; , \qquad \\
& = & 3 H^2 \Bigl\{1 - .6551 \times \frac{G H^2}{\pi} \Bigr\} \; .
\end{eqnarray}
This is insignificant even for the highest scale inflation ($G H^2 \ltwid 
10^{-12}$) consistent with the normalized CMB quadrupole and with the 
current upper bound on the scalar-to-tensor ratio \cite{WMAP}. The sign of 
the effect is nevertheless intriguing.

Because the field strength bilinear involves coincident passive fields it
requires renormalization even at leading logarithm order. This is evident 
from the factor of $\Gamma(2-\frac{D}2)$ in (\ref{Fexp}). As might be
expected, the divergence can be absorbed with terms proportional to
$\varphi^* \varphi$ and $(\varphi^* \varphi)^2$,
\begin{equation}
\Bigl( F_{\mu\nu} F_{\rho\sigma} \Bigr)_{\rm div} \equiv
\frac{H^D}{(4\pi)^{\frac{D}2}} \Gamma\Bigl(-\frac{D}2\Bigr) \, 4 z (1 + z) 
\Bigl( g_{\mu\rho} g_{\nu \sigma} - g_{\mu\sigma} g_{\nu\rho}\Bigr) \; .
\end{equation}
With this definition of the divergent part, the remaining finite part
has the following $D=4$ limit,
\begin{eqnarray}
\lefteqn{\Bigl( F_{\mu\nu} F_{\rho\sigma} \Bigr)_{\rm fin} \equiv 
\frac{H^4}{8 \pi^2} \Biggl\{-z - z (1 + z) } \nonumber \\
& & \hspace{.5cm} \times \Biggl[\psi\Bigl(\frac32 - \frac12 \sqrt{1- 8 z}\Bigr)
+ \psi\Bigl(\frac32 + \frac12 \sqrt{1 - 8 z} \Bigr)\Biggr] \Biggr\} \Bigl( 
g_{\mu\rho} g_{\nu \sigma} - g_{\mu\sigma} g_{\nu\rho}\Bigr) \; . 
\qquad \label{Ffin}
\end{eqnarray}
The late time limit of this quantity is strongly negative,
\begin{equation}
\lim_{t \rightarrow \infty} \Bigl\langle \Bigl( F_{\mu\nu} F_{\rho\sigma}
\Bigr) \Bigr\rangle \approx \frac{3 H^4}{8 \pi^2} \times -\frac{20.6305}{
2.16603} \times \Bigl(g_{\mu\rho} g_{\nu\sigma} - g_{\mu\sigma} g_{\nu\rho}
\Bigr) \; .
\end{equation}

Of course the quadratic and quartic parts of (\ref{Ffin}) can be adjusted
with the renormalization condition but the negative sign of the large field 
limiting form is unambiguous,
\begin{equation}
\Bigl( F_{\mu\nu} F_{\rho\sigma} \Bigr)_{\rm fin} \longrightarrow
\frac{H^2}{8 \pi^2} \Bigl[-z (z + 1) \ln(2 z + 2) - \frac13 z + 
O(1)\Bigr] \Bigl( g_{\mu\rho} g_{\nu \sigma} - g_{\mu\sigma} g_{\nu\rho}\Bigr) 
\; . \qquad 
\end{equation}
This indicates that the inflationary production of charged scalars increases
the electric field strength (for example, $\mu = \rho = 0$ and $\nu = \sigma 
= i$), relative to its vacuum value, while the magnetic field strength
(for example, $\mu = \rho = i$ and $\nu = \sigma = j$) is decreased. This
makes good physical sense. Although the average charge is zero, there is about
about one infrared scalar in each Hubble value. The local electric field is
necessarily dominated by this charge, so the square of the electric field
strength should increase. On the other hand, the nonzero scalar field 
strength engenders a positive photon mass which drives down the magnetic 
field by the Meissner effect. From expression (\ref{Fcon}) we see that the 
net electromagnetic contribution to the stress tensor is that of a negative 
cosmological constant at leading logarithm order. This indicates that the
damping of vacuum fluctuations in the magnetic field is greater than the
enhancement of the electric field.

\section{Epilogue}

Infrared logarithms are the manifestation of enhanced quantum effects 
mediated by massless, minimally coupled scalars and gravitons. We call 
these {\it active fields}. The continued growth of infrared logarithms 
must eventually overwhelm even the smallest loop-counting parameter. At 
this point perturbation theory breaks down and one must employ some sort 
of nonperturbative technique to follow the subsequent evolution. A 
reasonable approach is to sum the series of leading infrared logarithms.
Starobinski\u{\i} has developed a simple stochastic formalism which
accomplishes this for any model of purely active fields with nonderivative
interactions \cite{AAS,SY}.

More general models possess two sorts of complications: derivative
interactions and couplings with {\it passive fields}, that is, fields which
cannot cause infrared logarithms. We still do not have a general technique 
for handling derivative interactions. One deals with passive fields by 
integrating them out and then stochastically simplifying the resulting 
effective action of active fields. This amounts to computing the effective 
potential.  The theory then reduces to the form that Starobinski\u{\i} has 
already solved. This reduction was previously accomplished for Yukawa theory
\cite{MW1}, and we have done it here for SQED.

Note that one must integrate out passive fields from the VEV of any
operator. This can result in ultraviolet divergences even at leading
log order, as we found for both the field strength and the scalar kinetic
bilinears. The reason for this is that passive fields contribute factors 
of order one which multiply the infrared logarithms contributed by 
active fields. Whereas infrared logarithms derive entirely from the
long wavelength part of the free field mode sum, the factors of order one 
come as much from the ultraviolet as from the infrared. 

The nontrivial role of the ultraviolet also shows up in the fact that
certain renormalization counterterms can make leading order contributions.
In both Yukawa theory \cite{MW1} and SQED the one loop conformal and quartic
counterterms contribute. However, no other counterterms contribute at leading 
logarithm order, nor do any higher loop counterterms matter.

We have obtained nonperturbative results for the VEV's of $\varphi^*
\varphi$, $F_{\mu\nu} F_{\rho\sigma}$ and $T_{\mu\nu}$. Table~\ref{lims}
summarizes these. Our result is that the scalar approaches a
nonperturbatively large field strength. This confirms the conjecture of 
Davis, Dimopoulos, Prokopec and T\"ornkvist \cite{DDPT,DPTD}. The scalar 
remains perturbatively light, which means the computation is self-consistent.

Our result for the stress tensor is curious in two ways. First, although it 
takes the form $T_{\mu\nu} \longrightarrow -g_{\mu\nu} V_s$, the stress 
tensor potential $V_{\rm s}$ is not quite equal to the effective potential. 
This does not mean one potential is ``right'' and the other ``wrong.''
Rather, they are both the correct answers to slightly different physical
questions. The effective potential controls how the scalar evolves at 
leading logarithm order, whereas the stress tensor potential describes how
this evolution serves as a source for gravity. 

As shown in section 5, the difference between $V_{\rm s}$ and 
$V_{\rm eff}$ arises because almost all the factors of $H^2$ in $V_{\rm 
eff}$ are actually $R/12$ for a general metric. One consequence is that 
$V_{\rm eff}$ represents a peculiar modified gravity theory which may have 
important implications for cosmology. Lagrangians of the form $F(R)$ seem to be 
theoretically viable, and they can be tuned to give any desired evolution 
for the scale factor \cite{RPW3}. However, there is no justification for
such models from fundamental theory. In contrast, the modified gravity model
we get is uniquely fixed and thoroughly justified --- although it may not,
of course, do anything interesting.

The second peculiar thing about our result for the stress tensor is that
it reduces the vacuum energy. The physical interpretation for this may be
that the inflationary production of charged scalars polarizes the vacuum,
which lowers the energy of charged particles in this medium provided the
charge density is not too large. Supporting this conjecture is the fact that
the electromagnetic contribution to the stress tensor is that of a negative 
cosmological constant.

\newpage

\centerline{\bf Acknowledgements}

This work was partially supported by the Institute for Theoretical Physics
of Utrecht University, by the European Social fund and National resources 
$\Upsilon\Pi{\rm E}\Pi\Theta$-PythagorasII-2103, by European Union grants 
MRTN-CT-2004-512194 and FP-6-12679, by NSF grant PHY-0244714, and by the 
Institute for Fundamental Theory at the University of Florida.


\begin{thebibliography}{99}

\bibitem{TW3} N. C. Tsamis and R. P. Woodard, Class. Quant. Grav. {\bf 11}
(1994) 2969.

\bibitem{TW5} N. C. Tsamis and R. P. Woodard, Nucl. Phys. {\bf B474} (1996)
235, hep-ph/9602315.

\bibitem{RPW1} R. P. Woodard, ``Quantum Effects during Inflation,'' in
{\it Norman 2003, Quantum field theory under the influence of external
conditions} (Rinton Press, Princeton, 2004) ed. K. A. Milton, pp. 325-330,
astro-ph/0310757.

\bibitem{OW1} V. K. Onemli and R. P. Woodard, Class. Quant. Grav. {\bf 19}
(2002) 4607, gr-qc/0204065.

\bibitem{OW2} V. K. Onemli and R. P. Woodard, Phys. Rev. {\bf D70} (2004) 
107301, gr-qc/0406098.

\bibitem{VF} A. Vilenkin and L. H. Ford, Phys. Rev. {\bf D26} (1982) 1231.

\bibitem{L} A. D. Linde, Phys. Lett. {\bf B116} (1982) 335.

\bibitem{S} A. A. Starobinski\u{\i}, Phys. Lett. {\bf B117} (1982) 175.

\bibitem{BOW} T. Brunier, V. K. Onemli and R. P. Woodard, Class. Quant. Grav. 
{\bf 22} (2005) 59, gr-qc/0408080.

\bibitem{KO} E. O. Kahya and V. K. Onemli, Phys. Rev. {\bf D76} (2007) 
043512, gr-qc/0612026.

\bibitem{PTW} T. Prokopec, O. Tornkvist and R. P. Woodard, Ann. Phys.
{\bf 303} (2003) 251, gr-qc/0205130.

\bibitem{PW1} T. Prokopec and R. P. Woodard, Ann. Phys. {\bf 312} (2004) 1, 
gr-qc/0310056.

\bibitem{PTsW} T. Prokopec, N.C. Tsamis and R. P. Woodard, Class. Quant.
Grav. {\bf 24} (2007) 201, gr-qc/0607094.

\bibitem{PW2} T. Prokopec and R. P. Woodard, JHEP {\bf 0310} (2003) 059, 
astro-ph/0309593.

\bibitem{GP} B. Garbrecht and T. Prokopec, Phys. Rev. {\bf D73} (2006)
064036, gr-qc/0602011.

\bibitem{MW1} S. P. Miao and R. P. Woodard, Phys. Rev. {\bf D74} (2006)
044019, gr-qc/0602110.

\bibitem{TW0} N. C. Tsamis and R. P. Woodard, Phys. Rev. {\bf D54} (1996) 
2621, hep-ph/9602317.

\bibitem{TW1} N. C. Tsamis and R. P. Woodard, Ann. Phys. {\bf 253} (1997) 1,
hep-ph/9602316.

\bibitem{MW2} S. P. Miao and R. P. Woodard, Class. Quant. Grav. {\bf 23}
(2006) 1721, gr-qc/0511140.

\bibitem{MW3} S. P. Miao and R. P. Woodard, Phys. Rev. {\bf D74} (2006)
024021, gr-qc/0603135.

\bibitem{SW1} S. Weinberg, Phys. Rev. {\bf D72} (2005) 043514, hep-th/0506236.

\bibitem{BSV1} D. Boyanovsky, H. J. de Vega and N. G. Sanchez, Nucl. Phys. 
{\bf B747} (2006) 25, astro-ph/0503669.

\bibitem{BSV2} D. Boyanovsky, H. J. de Vega and N. G. Sanchez, Phys. Rev.
{\bf D72} (2005) 103006, astro-ph/0507596.

\bibitem{MS} M. Sloth, Nucl. Phys. {\bf B748} (2006) 149, astro-ph/0604488.

\bibitem{KC} K. Chaicherdsakul, Phys. Rev. {\bf D75} (2007) 063522,
hep-th/0611352.

\bibitem{BP} A. Biland\v{z}i\'c and T. Prokopec, Phys. Rev. {\bf D76}
(2007) 103507, arXiv:\-0704.1905 [astro-ph].

\bibitem{SW2} S. Weinberg, Phys. Rev. {\bf D74} (2006) 023508, hep-th/0605244.

\bibitem{CM} F. Cooper and E. Mottola, Phys. Rev. {\bf D36} (1987) 3114.

\bibitem{BCVHSS} D. Boyanovsky, D. Cormier, H. J. de Vega, R. Holman, A.
Singh and M. Srednicki, Phys. Rev. {\bf D56} (1997) 1939, hep-ph/9703327.

\bibitem{AAS} A. A. Starobinski\u{\i}, ``Stochastic de Sitter (inflationary)
stage in the early universe,'' in {\it Field Theory, Quantum Gravity and
Strings}, ed. H. J. de Vega and N. Sanchez (Springer-Verlag, Berlin, 1986)
pp. 107-126.

\bibitem{AV} A. Vilenkin, Phys. Rev. {\bf D27} (1983) 2848.

\bibitem{NS} Y. Nambu and M. Sasaki, Phys. Lett. {\bf 219} (1989) 240.

\bibitem{GLM} A. S. Goncharov, A. D. Linde and V. F. Mukhanov, Int. J.
Mod. Phys. {\bf A2} (1987) 561.

\bibitem{LM} A. D. Linde and A. Mezhlumian, Phys. Lett. {\bf B307} (1993)
25, gr-qc/9304015.

\bibitem{SJR} S. J. Rey, Nucl. Phys. {\bf B284} (1987) 706.

\bibitem{SNN} M. Sasaki, Y. Nambu and K. I. Nakao, Nucl. Phys. {\bf B308}
(1988) 868.

\bibitem{WV} S. Winitzki and A. Vilenkin, Phys. Rev. {\bf D61} (2000)
084008, gr-qc/9911029.

\bibitem{MM} J. Martin and M. Musso, Phys. Rev. {\bf D73} (2006) 043517,
hep-th/0511292.

\bibitem{SY} A. A. Starobinski\u{\i} and J. Yokoyama, Phys. Rev. {\bf D50}
(1994) 6357, astro-ph/9407016.

\bibitem{RPW2} R. P. Woodard, Nucl. Phys. Proc. Suppl. {\bf 148} (2005) 108,
astro-ph/0502556.

\bibitem{TW2} N. C. Tsamis and R. P. Woodard, Nucl. Phys. {\bf B724} (2005)
295, gr-qc/0505115.

\bibitem{DDPT} A. C. Davis, K. Dimopoulos, T. Prokopec and O. T\"ornkvist,
Phys. Lett. {\bf B501} (2001) 165, astro-ph/0007214.

\bibitem{DPTD} K. Dimopoulos, T. Prokopec, O. T\"ornkvist and A. C. Davis,
Phys. Rev. {\bf D65} (2002) 063505, astro-ph/0108093.

\bibitem{YF} C. N. Yang and D. Feldman, Phys. Rev. \textbf{79} (1950) 972.

\bibitem{FP} L. H. Ford and L. Parker, Phys. Rev. {\bf D16} (1977) 245.

\bibitem{AV2} A. Vilenkin, Nucl. Phys. {\bf B226} (1983) 527.

\bibitem{Mus} M. Musso, ``A New diagrammatic representation for correlation
functions in the in-in formalism,'' hep-th/0611258.

\bibitem{GR} I. S. Gradshteyn and I. M. Ryzhik, {\it Table of Integrals,
Series and Products,} 4th ed. (Academic Press, New York, 1965).

\bibitem{ALV} L. Accardi, Y. G. Lu and I. Volovich, {\it Quantum Theory and
Its Stochastic Limit} (Springer-Verlag, Berlin, 2002).

\bibitem{KW} E. O. Kahya and R. P. Woodard, Phys. Rev. {\bf D72} (2005)
104001, gr-qc/0508015.

\bibitem{RJ} R. D. Jordan, Phys. Rev. {\bf D33} (1986) 444.

\bibitem{PCDJR} P. Candelas and D. J. Raine, Phys. Rev. {\bf D12} (1975) 965.

\bibitem{JSDRC} J. S. Dowker and R. Critchley, Phys. Rev. {\bf D13} (1976)

\bibitem{SRCEW} S. R. Coleman and E. Weinberg, Phys. Rev. {\bf D7} (1973)
1888.

\bibitem{Jackiw} R. Jackiw, Phys. Rev. {\bf D9} (1974) 1686.

\bibitem{TW4} N. C. Tsamis and R. P. Woodard, J. Math. Phys. {\bf 48} (2007)
052306, gr-qc/0608069.

\bibitem{KW2} E. O. Kahya and R. P. Woodard, Phys. Rev. {\bf D74} (2006)
084012, gr-qc/0608049.

\bibitem{Gar} B. Garbrecht, Nucl. Phys. {\bf B784} (2007) 118,
hep-ph/0612011.

\bibitem{Wolfram} S. Wolfram, {\it The Mathematica Book, Third Edition}
(Cambridge University Press, 1996).

\bibitem{PTW2} T. Prokopec, O. T\"ornkvist and R. P. Woodard, Phys. Rev.
Lett. {\bf 89} (2002) 101301, astro-ph/0205331.

\bibitem{PW3} T. Prokopec and R. P. Woodard, Am. J. Phys. {\bf 72} (2004) 60,
astro-ph/0303358.

\bibitem{PP1} T. Prokopec and E. Puchwein, JCAP {\bf 0404} (2004) 007,
astro-ph/\-0312274.

\bibitem{PP2} T. Prokopec and E. Puchwein, Phys. Rev. {\bf D70} (2004) 
043004, astro-ph/0403335.

\bibitem{WMAP} D. N. Spergel et al., Astrophys. J. Suppl. {\bf 170}
(2007) 377, astro-ph/0603449.

\bibitem{RPW3} R. P. Woodard, Lect. Notes Phys. {\bf 720} (2007) 403,
astro-ph/0601672.

\end{thebibliography}
\end{document}